\documentclass[fleqn,usenatbib]{mnras}

\usepackage{newtxtext,newtxmath}

\usepackage[T1]{fontenc}
\usepackage{ae,aecompl}

\usepackage{graphicx}	
\usepackage{amsmath}	
\usepackage{amssymb}	
\usepackage[dvipsnames]{xcolor}

\title [Magnetic braking and ELM WDs]{Convection and rotation boosted prescription of magnetic braking: application to the formation of extremely low-mass white dwarfs}

\author[L. T. T. Soethe and S. O. Kepler]{
L. T. T. Soethe$^{1}$\thanks{E-mail: tayno32@gmail.com (LTTS)}
and S. O. Kepler$^{1}$
\\
$^{1}$Instituto de F\'{i}sica, Universidade Federal do Rio Grande do Sul, Av. Bento Gon\c{c}alves 9500, Porto Alegre 91501-970, RS, Brazil\\
}

\date{Accepted XXX. Received YYY; in original form ZZZ}

\pubyear{2021}

\begin{document}
\label{firstpage}
\pagerange{\pageref{firstpage}--\pageref{lastpage}}
\maketitle


\begin{abstract}
Extremely low-mass white dwarfs (ELM WDs) are the result of binary evolution in which a low-mass donor star is stripped by its companion leaving behind a helium-core white dwarf.
We explore the formation of ELM WDs in binary systems considering the Convection And Rotation Boosted magnetic braking treatment.
Our evolutionary sequences were calculated using the MESA code, with initial masses of 1.0 and 1.2~$M_{\odot}$ (donor), and 1.4 (accretor), compatible with low mass X-ray binaries (LMXB) systems.
We obtain ELM models in the range $0.15$ to $0.27~M_{\odot}$ from a broad range of initial orbital periods, 1 to 25~d.
The bifurcation period, where the initial period is equal to the final period, ranges from 20 to 25~days.
In addition to LMXBs, we show that ultra-compact X-ray binaries (UCXB) and wide-orbit binary millisecond pulsars can also be formed.
The relation between mass and orbital period obtained is compatible with the observational data from He white dwarf companions to pulsars.
\end{abstract}




\section{Introduction}
\label{sec:Intro}
\defcitealias{Rappaport1983MB}{R83}
\defcitealias{Van2019}{V19}
\defcitealias{Van2018}{V18}

Extremely low-mass (ELM) white dwarfs are helium-core white dwarfs with masses $M \leqslant 0.3~M_{\odot}$, most likely formed through binary interactions,
since they would take more than the age of the Universe to evolve out of the main sequence as single stars.
There are several definitions for the upper mass limit for the class in the literature, e.g.,
$0.18~M_{\odot}$ \citep{Sun2018Formation},
$0.20~M_{\odot}$ \citep{Kawka2009,Althaus2013,Chen2017},
$0.25~M_{\odot}$ \citep{Hermes2013mar,Hermes2013dec},
$0.30~M_{\odot}$ \citep{LiChenHan2019,Pelisoli2019GaiaELM}.
In the most common binary evolution scenario, a main sequence star reaches the red giant branch, fills its Roche-lobe, and then starts stable mass transfer through Roche-lobe overflow (RLOF) to the accretor companion --- either a WD
\citep{Kilic2007,Kulkarni2010,TaurisLangerKramer2012,Sun2018Formation,LiChenHan2019}
or a neutron star
\citep[NS;][]{BhattacharyaHeuvel1991,Podsiadlowski2002,Kerkwijk2005,Shao2015,Cadelano2019,Sanchez2020}.
These systems are then associated with the cataclysmic variables (CV)
and low-mass X-ray binary (LMXB) systems, respectively.

Most known ELMs will merge in less than a Hubble time resulting in new exotic objects such as R Corona Borealis stars \citep{Webbink1984,ZhangJefferyChenHan2014}, underluminous supernovae \citep{BildstenShen2007,BrownKilicPrietoKenyon2011},
type Ia supernovae \citep{IbenTutukov1984}, and AM CVn systems \citep{BreedtGansickeMarsh2012,BrownKilicKenyonGianninas2016}.
The shortest period ELM WD binaries will serve as multi messenger laboratories \citep{Korol2017,Kupfer2018}. In particular, \citet{Brown2020-1201s}
discovered the first He$+$He WD LISA verification binary, a dominant LISA source along with He-CO double white dwarfs \citep{Lamberts2019}.
Furthermore, it is expected that compact post-LMXB/pre-UCXB systems can provide a very accurate measurement for the mass of neutron stars, imposing restrictions on their state equation \citep{Tauris2018}.

In the past decade, more than a hundred systems containing ELMs have been discovered by several surveys. For example, the ELM Survey \citep{ELMSurveyI,ELMSurveyII,ELMSurveyIII,ELMSurveyIV,ELMSurveyV,ELMSurveyVI,ELMSurveyVII,Brown2020ELMVIII} now counts 62 ELMs in the clean sample\footnote{They define the clean ELM WD sample as: ELM WDs in the dereddened magnitude range $15<g_{0}<20$, located in the SDSS footprint, with $8800<T_\text{eff}/\text{K}<22,000$ and $5.5 \leqslant \log (g) \leqslant 7.1$.}, where 65\% were identified as galactic disk objects and 35\% as halo objects at distances up to 3.752~kpc. Orbital periods were measured as $12.8\,\mathrm{min} \leq P \leq 1.48567\,\mathrm{d}$ and estimated ELMs masses are in the range $0.15 \leq M/M_{\odot} \leq 0.30$. In the ELM Survey, mass estimates are obtained spectroscopically by fitting $\log (g)$ and $T_\text{eff}$ to a set of atmospheric models, without a proper accounting of metallicity difference or envelope mass differences. Therefore, these masses are not precise estimates. The ELM Survey has started the search for ELM WDs in the southern sky \citep{Kosakowski2020ELMSSI}, identifying ELMs in systems with periods as short as 2~h and also the second closest (72~pc) known ELM to date\footnote{The nearest ELM WD position was recently overtaken by an $\approx 0.17~M_{\odot}$ ELM and $\approx 71~\text{pc}$ away \citep{Kawka2020}.}.

\citet{Chen2017,Sun2018Formation} concluded that the formation of an ELM white dwarf with $M \lesssim 0.18\text{--}0.20~M_{\odot}$ by unstable mass transfer or a common-envelope (CE) event is unlikely.
Also, \citet{LiChenHan2019} found that ELM WDs with $M \lesssim 0.3~M_{\odot}$ in double degenerate systems may be formed either from a stable mass transfer process or common-envelope ejection, although the Roche-lobe formation channel has a greater contribution to the formation of He WD with mass $\lesssim 0.22~M_{\odot}$, and the common envelope channel for higher masses.
The currently-observed binarity rate of known ELM WDs --- close to 100 per cent --- supports both channels. Their companions include millisecond pulsars, main sequence stars \citep[the so-called EL CVns,][]{Maxted2014,Chen2017},
hot subdwarf B stars \citep{Kupfer2015},
and more commonly canonical mass WDs \citep[e.g.,][]{Kilic2007,Kulkarni2010,TaurisLangerKramer2012,BrownKilicGianninas2017}.
However, observations are likely to be biased considering current optical spectroscopy is more sensitive to short orbital period systems \citep{ELMSurveyI,ELMSurveyIV,Toloza2019}.
In addition, failure to observe isolated ELM WDs does not necessarily mean that they do not exist.

The formation and evolution of ELM WDs through the LMXB and CV channels was studied extensively in the literature, although the input physics
 --- e.g., donor and accretor masses, accretion efficiency, metallicity, use of rotation and diffusion, mechanisms of angular momentum loss, mass transfer formalism ---
considered in each of them is widely varied
\citep[e.g.,][]{Podsiadlowski2002,Serenelli2002,Panei2007,Althaus2009,Lin2011,Althaus2013,Istrate2014,Istrate2016Models-ads}.

In general, theoretical models are able to reproduce with good accuracy the physical parameters observed in ELM WDs, such as chemical abundance, final mass, effective temperature, and surface gravity.
Despite these successes, there are still open questions.
For instance, a severe fine-tuning --- of the order of a dozen minutes --- in the initial orbital period
was necessary to reproduce the observed millisecond pulsars in compact (2~h $\leq P \leq$ 9~h)
binaries with He WD companions of mass $\lesssim 0.20~M_{\odot}$ \citep{Istrate2014}.
This extreme fine-tuning in the initial orbital period suggests that something is missing in the standard input physics of LMXB modelling.
As most angular momentum loss mechanisms are reasonably well understood, the problem seems to fall on the magnetic braking.
In addition, irreconcilable discrepancies are found in some systems where it is possible to obtain the mass of the components with good precision and independently of the evolutionary ELM WDs models \citep{Sanchez2020,LiuIstrate2020}.

The most used empirical torque formula for magnetic braking of
\citet[][hereafter \citetalias{Rappaport1983MB}]{Rappaport1983MB} was derived from observations of solar-mass main-sequence stars that exhibit a strong correlation between equatorial rotation velocity and age
\citep{Skumanich1972,Smith1979}.
\citetalias{Rappaport1983MB} formula has been used and discussed extensively in the literature in calculations of evolution and formation of binary systems, including low and intermediate-mass X-ray binaries, millisecond radio pulsars, cataclysmic binaries, and subdwarf B stars \citep{Patterson1984, BhattacharyaHeuvel1991, Podsiadlowski2002, Han2003sdB, Knigge2011}.
\citet[][hereafter \citetalias{Van2018}]{Van2018}
and
\citet[][hereafter \citetalias{Van2019}]{Van2019}
proposed a physically motivated magnetic braking prescription.
In their formalism, the magnetic braking of the system is calculated from a rotating spherically symmetric star,
considering a radial magnetic field in the dipole approximation. 
The mass lost by winds --- assumed isotropic --- corotates with the star up to a distance that depends on the size of the convective zone, rotation velocity, and the magnitude of the surface magnetic field. 
\citetalias{Van2018} showed that the braking law from \citetalias{Rappaport1983MB} is not suitable to explain most of the observed persistent LMXBs, specially the observed mass transfer rates.
Instead, the \citetalias{Van2019} prescription was successful in reproducing the observed mass transfer rates of persistent LMXB for all observed mass ratio and orbital periods.
Furthermore, \citet{Deng2021} studied the LMXB evolution with five proposed magnetic braking laws and found that both the \citetalias{Van2018} and the \citetalias{Van2019} laws are more preferred in reproducing the properties of persistent and transient LMXBs systems.

Although the formulations \citetalias{Van2018} and \citetalias{Van2019} have very similar physical motivations, the formula presented in \citetalias{Van2018} has three free parameters. This makes the results depend on the chosen parameters.
On the other hand, the \citetalias{Van2019} formulation has a more consistent deduction and does not have any free parameters.

\citet{ChenTauris2021} used the \citetalias{Van2018} formulation to study the evolutionary link from low-mass X-ray binaries (LMXBs) to binary millisecond pulsars (BMSPs) and ultra-compact X-ray binaries (UCXBs).
Although \citet{ChenTauris2021} focused on the parameter space for the formation of UCXBs, they found that the \citetalias{Van2018} prescription fails to form wide-orbit BMSPs.
In addition, both \citetalias{Van2018} and \citet{ChenTauris2021} made it clear that the issue of free parameters gave significantly different outcomes, making it more difficult to draw general conclusions.

In this work we apply the Convection And
Rotation Boosted (CARB) prescription for the magnetic braking presented by \citetalias{Van2019} to study the formation and evolution of low-mass and ELM WDs.
We show that the use of CARB magnetic braking reproduces the LMXB phase as well as being able to form UCXBs and wide BMSPs systems.

The layout of this paper is as follows.
In Section~\ref{sec:NumMetSim} we describe the physical ingredients
considered in computing the grid of models;
in Section~\ref{sec:Results} we present the results obtained from the fully evolutionary computations and compare with observational data.
Concluding remarks are presented in Section~\ref{sec:Conclusions}.

\section{Numerical Methods and Simulations}
\label{sec:NumMetSim}

The model grid presented in this work is computed using Modules for Experiments in Stellar Astrophysics code
\citep[\textsc{mesa}][]{Paxton2011, Paxton2013, Paxton2015, Paxton2018, Paxton2019}, release 11701.
We compute the binary evolution of the system following the evolution of the donor star from the zero age main sequence (ZAMS) until it reached a model age of 14~Gyr.
The accretor is treated as a point mass.
For the rest of the manuscript, we adopt the nomenclature `d' for the donor and `a' for the accretor.
Initial and final ages will be indicated by `i' and `f', respectively.

Below we describe the input physics and the computational details used to calculate the evolutionary sequences.

\subsection{Stellar evolution input}
\label{subsec:DonorS}

The equation of state (EoS) is a blend of the OPAL \citep{Rogers2002}, SCVH \citep{Saumon1995}, PTEH \citep{Pols1995}, HELM
\citep{Timmes2000}, and PC \citep{Potekhin2010} EoSs. A smooth transition between the EoSs guarantees the appropriate usage across the entire required range of density and temperature.
Radiative opacities are primarily from OPAL \citep{Iglesias1993, Iglesias1996}, with low--temperature data from \citet{Ferguson2005} and the high--temperature, Compton-scattering dominated regime by \citet{Buchler1976}. Electron conduction opacities are from \citet{Cassisi2007}.

Nuclear reaction rates are from JINA REACLIB \citep{Cyburt2010} plus additional weak reaction rates \citep{Fuller1985, Oda1994, Langanke2000}.
Screening is included via the prescription of \citet{Chugunov2007}. Thermal neutrino loss rates are from \citet{Itoh1996a}. Hydrogen burning (p-p chain and CNO cycle) are computed by using the \texttt{cno\textunderscore extras.net} network that accounts for the following 21 isotopes:
${}^{1}$H, ${}^{3}$He, ${}^{4}$He, ${}^{12}$C, ${}^{13}$C, ${}^{13}$N, ${}^{14}$N, ${}^{15}$N, ${}^{14}$O, ${}^{15}$O, ${}^{16}$O, ${}^{17}$O, ${}^{18}$O, ${}^{17}$F, ${}^{18}$F, ${}^{19}$F, ${}^{18}$Ne, ${}^{19}$Ne, ${}^{20}$Ne, ${}^{22}$Mg and ${}^{24}$Mg.
As calcium is one of the easiest elements to be detected in the spectra of a WD, we also include the ${}^{40}$Ca isotope in our models.

Convection is treated using the simple local formulation of the mixing$-$length theory \citep{Bohm1958MLT} in the variation of \citet{Henyey1965MLT} allowing the convective efficiency to vary with the opacity. Following \citet{Istrate2016Models-ads}, $\alpha_{\text{MLT}}=2$ is adopted as the mixing length parameter. We consider the Ledoux criterion of stability, which takes into account the influence of composition gradients on mixing. Semiconvection is considered in the regions unstable with respect to the Schwarzschild criterion but stable to Ledoux, with an efficiency parameter $\alpha_{\text{sc}}=0.001$.
Thermohaline mixing is included during all the evolution with efficiency $\alpha_{\text{th}}=1$.

Following \citet{Istrate2016Models-ads}, for each burning/non-burning core/shell region we include an exponential overshooting below and above the interface limit with
$f=0.01$ and $f_{0}=0.005$.
An extra step overshooting of $f=0.25$ and $f_{0}=0.05$
above the burning H core is also included.

For comparison, we calculated models with and without diffusion. For the models with diffusion, element diffusion and gravitational settling is included by solving Burger's equations and using the method of \citet{Thoul1994} with diffusion coefficients similar to \citet{Iben1985}.
Unlike \citet{Istrate2016Models-ads} who clumps elements into classes, we treat each isotope as a separate class. This treatment does not affect the results, as we will discuss in Section~\ref{sec:Results}.

The effects of rotation may become important near the rotation limit and are included by following \citet{Heger2000} and \citet{Heger2005}, where we include the effects for four different rotationally induced mixing processes: Goldreich-Schubert-Fricke instability, Eddington-Sweet circulation, secular shear instability and dynamical shear instability.
More details about rotation and rotational element transport mechanisms can be found in the review of \citet{SalarisCassisi2017}.
The Spruit-Tayler dynamo transports angular momentum and chemicals by magnetic fields. Here we must set two efficiency factors to calibrate the diffusion coefficients: the contribution of the rotationally induced instabilities to the diffusion coefficient is reduced by the factor $f_{\text{c}}=1/30$, and the sensitivity of the rotationally induced mixing is $f_{\mu}=0.05$.
These values follow \citet{Istrate2016Models-ads}. See \citet{Heger2000} for a discussion of these calibration parameters.

For the atmosphere boundary conditions we consider the simple photosphere option for the pre-WD phase and the hydrogen atmosphere tables for cool white dwarfs from \citet{Rohrmann2011} for $T_{\text{eff}}<10,000~\text{K}$ and $\text{log}_{10}(L/L_{\odot})<-2$.

\subsection{Binary evolution input}
Given the initial donor ($M_\text{d,i}$) and accretor ($M_\text{a,i}$) masses, each system starts with both stars
in a circular orbit with separation $a$, and initial orbital period $P_\text{i}=2\pi \left[a^3/G(M_\text{d,i}+M_\text{a,i})\right]^\frac{1}{2}$.
At the beginning of evolution, the rotation of the donor star is relaxed to the orbital period of the system by applying tidal torque considering the synchronisation
timescale for convective envelopes \citep{Hurley2002}.

The rate of change of the angular momentum of the system is computed considering contributions from gravitational wave radiation, mass loss, magnetic braking, and spin$-$orbit coupling as follow:
\begin{equation}
    \dot{J}_{\text{orb}}=\dot{J}_{\text{gr}}+\dot{J}_{\text{ml}}+\dot{J}_{\text{mb}}+\dot{J}_{\text{ls}}.
	\label{eq:Jdot}
\end{equation}

In the absence of outer convective zones,
the gravitational wave radiation term dominates in very compact orbits ($P \lesssim 3~\text{d}$) and is given by
\citep[see, e.g.,][]{Landau1975CTF}:
\begin{equation}
\dot{J}_{\text{gr}}= -\frac{32}{5c^{5}}
\left( \frac{2 \pi G}{P} \right)^{7/3}
\frac{(M_\text{d}M_\text{a})^{2}}{(M_\text{d}+M_\text{a})^{2/3}},
\label{eq:Jgw}
\end{equation}
where
$G$ is the gravitational constant and $c$ is the speed of light in vacuum.

The mass transfer stability criteria is given by
\citet{Soberman1997}.
We calculated models with $\beta_\text{mt}=0.3$ and 0.8 for the fraction of mass lost from the vicinity of the accretor as fast wind, implying a mass transfer efficiency of 70 and 20 per cent, respectively.
The former value is for better comparison with models in the literature, and the latter was chosen given that there is observational evidence that mass transfer during the LMXB phase is extremely inefficient, corresponding to accretion efficiencies of only $\sim 5\text{--}40$ per cent \citep{Antoniadis2012,Antoniadis2013,Antoniadis2016}.

Roche lobe radii in binary systems are computed using the fit of \citet{Eggleton1983}, while mass transfer rates are determined following the prescription of \citet{Kolb1990}.
We consider the RLOF to take place when the mass transfer rate exceeds the value of
$\dot{M} = 10^{-10} ~M_{\odot}~\text{yr}^{-1}$.
We stress that this is just an arbitrary limit that has no effect in our results.

\subsection{Magnetic braking prescription}

Magnetic braking plays an important role, specially in the early stages of evolution in interacting binary systems.
Considering that stars of different spectral types on the main sequence have different rotational speeds,
\citet{Schatzman1962} was the first to suggest that the convective envelope could be the reason for some stars to have low rotation velocities.
He suggested that in convective stars the high magnetic field forces that ejected matter spinning along with the star, even at very high distances, carries a large amount of angular momentum per unit mass.
The first numerical estimate came from 
\citet{Skumanich1972},
who showed that the equatorial rotation velocities of G-type main sequence stars decrease with time,
suggesting the empirical dependence $\Omega \propto t^{-1/2}$.
A more elaborate expression appeared in the seminal work of
\citet{Rappaport1983MB},
where the mass and radius of the star are time dependent quantities.
The MESA implementation follows
\begin{equation}
\dot{J}_{\text{mb}}=-6.82\times10^{34}
\left( \frac{M_\text{d}}{M_{\odot}} \right)
\left( \frac{R_\text{d}}{R_{\odot}} \right)^{\gamma}
\left( \frac{1 \ \text{d}}{P} \right)^{3} ~ \text{,}
	\label{eq:Jmb}
\end{equation}
where $R_\text{d}$ is the radius of the donor and $\gamma=4$ is adopted for the magnetic braking index.

\citet{Van2019} alternative formulation for the magnetic braking is called the Convection And Rotation Boosted (CARB) magnetic braking.
CARB magnetic braking considers the dependence of the Alfv\'{e}n radius ($R_\text{A}$) on the rotation rate of the donor, and the dependence of the magnetic field strength on the outer convective zone.
The Alfv\'{e}n surface is the surface where the ram pressure is equal to the magnetic pressure \citep{MestelSpruit1987}, marking the maximum distance at which the stellar wind is still in corotation with the star. At larger distances, the mass is assumed to be lost from the star.
Spherical symmetry is assumed, which results in the angular momentum lost by magnetic breaking through an Alfv\'{e}n surface to be \citep{Weber1967,Mestel1968}
\begin{equation}
 \dot{J}_\text{mb}=
 -\frac{2}{3} \Omega \dot{M}_{W} R_{A}^{2} ~ \text{,}
 \label{eq:JmbWeber}
\end{equation}
where $\Omega$ is the rotation rate and $\dot{M}_{W}$ is the wind mass-loss rate.
The wind mass-loss is assumed to be isotropic, and depends on the density and the velocity of the mass flux through the Alfv\'{e}n surface.
Including the effects of rotation in the expression for the Alfv\'{e}n radius, the CARB magnetic braking then reads\footnote{We note that there is a missing minus factor in the exponential argument of the wind mass-loss rate in equation (5) of \citet{Van2019}. Also, in the code made available online by the authors, the argument 2/3 of the exponential in line 320 should be 1/3 in order to make it correctly fit in the corresponding equations in the paper.}
\begin{equation}
\begin{split}
 \dot{J}_\text{mb,CARB}&=
 -\frac{2}{3} \Omega \dot{M}_{W}^{-1/3} R_\text{d}^{14/3} \left( v_\text{esc}^{2} + \frac{2 \Omega R_\text{d}^{2}} {K_{2}^{2}} \right)^{-2/3} \\
 & \times\ \Omega_{\odot} B_{\odot}^{8/3} \left( \frac{\Omega}{\Omega_{\odot}} \right)^{11/3} \left( \frac{\tau_\text{conv}}{\tau_{\odot \text{,conv}}} \right)^{8/3}
~ \text{,}
\end{split}
\label{dotJmbCARB}
\end{equation}
where $v_\text{esc}= \sqrt{2GM/R}$ is the escape velocity, $K_{2}=0.07$ is a constant obtained via simulations \citep{Reville2015}.
$\tau_\text{conv}$ is the turnover time of convective eddies, given by
\begin{equation}
 \tau_\text{conv}=\int_{R_\text{bot}}^{R_\text{top}}
 \frac{\textrm{d}r}{v_\text{conv}} ~ \text{,}
 \label{eq:tauconv}
\end{equation}
where $v_\text{conv}$ is the local convective velocity 
and the integration limits, $R_\text{bot}$ and $R_\text{top}$, are the bottom and the top of the outer convective zone, respectively.
Thus, this description assumes the total magnetic field is generated by the convection eddies, i.e., no fossil fields.
The surface magnetic field is given by $B_\text{s}=\tau_\text{conv} \Omega$, as in \citet{Ivanova2006,Van2018}.
The last terms of equation~(\ref{dotJmbCARB}) is normalized according to a solar calibration, resulting in
$\Omega_{\odot} \approx 3 \times 10^{-6}~\text{s}^{-1}$ and
$\tau_{\odot \text{,conv}}=2.8 \times 10^{6}~\text{s}$.
We adopt the \citet{Reimers1975} wind mass-loss scheme, as it was done in \citet{Van2019}.
The prescriptions of \citet{Rappaport1983MB} and \citet{Van2019} differ essentially by the fact that the former is an empirical fit, and the latter is obtained through a self-consistent deduction considering wind mass loss, rotation, and that the magnetic field is generated due to motions in the convective zone.
The main limitations of this model will be addressed in the conclusion (Section~\ref{sec:Conclusions}).

\subsection{Model grid}
\label{subsec:Grid}

We consider initial donor masses of 1.0 and 1.2 $M_{\odot}$
with initial metallicity of $Z=0.02$.
The initial accretor mass is 1.4~$M_{\odot}$, consistent with neutron star companions.
Systems are initialised with orbital periods between one and 300 days.
The exact minimum initial orbital period for each configuration is defined so that the systems are completely detached at the end of the evolution.
The step in the initial orbital period varies between one and 100 days, for short and long initial orbital periods, respectively.
Rotation is considered in all configurations, and element diffusion is also included in some configurations,
as listed in Table~\ref{tab:setups}.
For brevity, the general discussion of the effect of the CARB magnetic braking on the formation of binary systems containing ELM white dwarfs will be done with a single initial configuration, facilitating the comparison with the models of \citet[][Sections~\ref{subsec:GeneralEffects} and \ref{subsec:PeriodBif}]{Istrate2016Models-ads}.
More detailed discussions on different initial masses, accretion efficiencies, and diffusion of elements will be presented on
Section~\ref{subsec:ELM+NS},
where we show the $P_\text{i}$--$P_\text{f}$ and the $M_\text{f}$--$P_\text{f}$ relations.
The main properties --- i.e., initial and final masses of both components, initial and final orbital period, number of hydrogen shell flashes --- of our model grid can be found in Appendix~\ref{sec:Appendix-CompMod}.
For quick use of our results, in Appendix~\ref{sec:Appendix-PolFits} we provide polynomial fits to the final ELM WD mass ($M_\text{d,f}$) as a function of the initial orbital period ($P_\text{i}$).
We also present a simple estimate of the gravitational wave strain and the merging time for our most compact binary models, as detailed in Appendix~\ref{sec:Appendix-GW}.

\begin{table}
	\centering
	\caption{Summary of the three different initial setups studied in this work. The second and third columns are the initial masses of the donor and the accretor, respectively. The fourth column is the initial metallicity. The fifth column is the fraction of mass lost from the vicinity of the accretor. The last column indicates whether the diffusion of elements was considered.
	}
	\label{tab:setups}
	\begin{tabular}{cccccc} 
		\hline
		\# & $M_\text{i,d}/M_{\odot}$ & $M_\text{i,a}/M_{\odot}$ & Z & $\beta_\text{mt}$ & rot/dif\\
		\hline
		1 & 1.0 & 1.4 & 0.02 & 0.3 & rot\\
		2 & 1.2 & 1.4 & 0.02 & 0.3 & rot\\
	    3 & 1.0 & 1.4 & 0.02 & 0.8 & rot+dif\\
		\hline
	\end{tabular}
\end{table}
%

\section{Results and discussion}
\label{sec:Results}

\subsection{Effects of the magnetic braking}
\label{subsec:GeneralEffects}

We first show how each of the two prescriptions for magnetic braking differ in the formation of LMXBs systems.
We start with one single initial setup and vary the initial orbital period searching for representative cases of different evolutionary scenarios.

\begin{figure} 
	\includegraphics[width=\columnwidth]{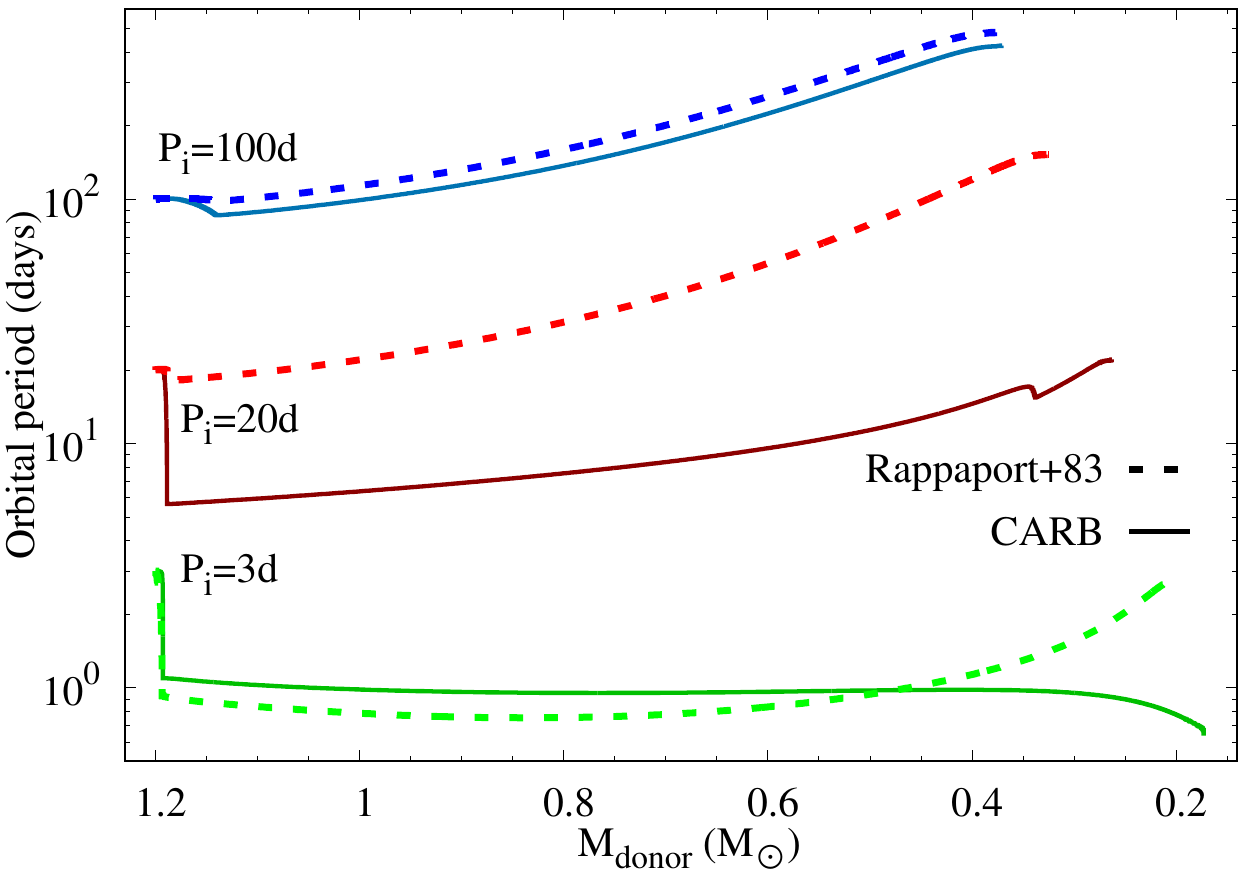}
    \caption{The evolution of orbital period as a function of decreasing donor resulting mass. Two prescriptions for magnetic braking are compared: \citet{Rappaport1983MB} (dashed lines) and \citet{Van2019} (solid lines). For each prescription, three initial orbital periods are analysed: 3 (green), 20 (red), and 100 (blue) days. Initial masses are $M_{\text{d,i}} = 1.2~M_{\odot}$ and $M_{\text{a,i}} = 1.4~M_{\odot}$ for all sequences.}
    \label{fig:plot-Md-P}
\end{figure}

\begin{figure} 
	\includegraphics[width=\columnwidth]{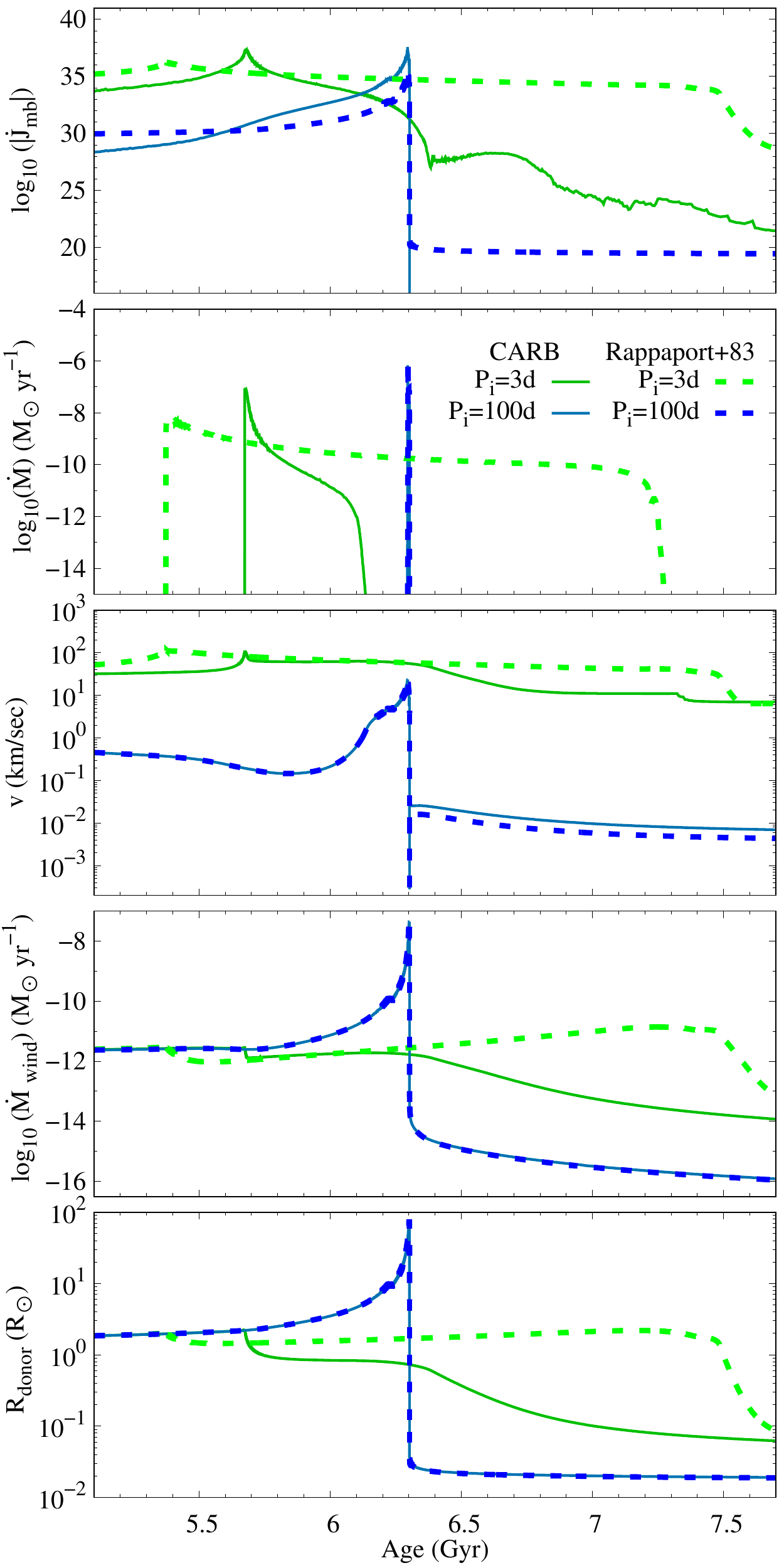}
    \caption{The evolution of key parameters during the mass transfer epoch. Two prescriptions for magnetic braking are compared: \citet[][dashed lines]{Rappaport1983MB} and \citet[][solid lines]{Van2019}.
    For each prescription, two initial orbital periods are analysed: 3 (green) and 100 (blue) days.
    Initial masses are $M_{\text{d,i}} = 1.2~M_{\odot}$ and $M_{\text{a,i}} = 1.4~M_{\odot}$ for all sequences.
    Magnetic braking (first panel); mass transfer rate (second panel); surface rotational velocity at equator (third panel); wind mass loss (fourth panel); and radius of the donor star (fifth panel) are shown.}
    \label{fig:plot-age-all5}
\end{figure}

Fig.~\ref{fig:plot-Md-P}
shows the
orbital period evolution as a function of the resulting donor mass for three systems, with
$M_\text{d,i}=1.2~M_{\odot}$ and $M_\text{a,i}=1.4~M_{\odot}$, and for three different initial orbital periods, $P_\text{i}=3$ (green), 20 (red), and 100 (blue) days.
Each configuration is shown for both \citetalias{Rappaport1983MB} (dashed lines) and CARB (\citetalias{Van2019}, solid lines) magnetic braking, totalizing six evolutionary sequences.
The evolution of the main physical quantities that govern magnetic braking are shown in Fig.~\ref{fig:plot-age-all5} for the cases of $P_\text{i} = 3$ and 100 days.
All other parameters are identical for all sequences.
All sequences produce detached He-core ELM white dwarf + neutron star binaries as output.


From short to long initial orbital periods, the final masses for
CARB \citepalias{Rappaport1983MB} sequences are 0.173 (0.205), 0.263 (0.325), and 0.371 (0.377) solar masses, respectively.
In the same way, the respectively final orbital periods are 0.64 (2.55), 22 (152), and 424 (481) days.

Fig.~\ref{fig:plot-Md-P} shows the existence of two different evolutionary scenarios when we compare the prescriptions of \citetalias{Rappaport1983MB} and CARB for magnetic braking.
The first scenario corresponds to intermediate and long initial orbital periods ($\sim$20--100~days), where the evolution of the donor mass star has similar shapes for the two prescriptions of magnetic braking.
The orbital period increases as the donor star loses mass, regardless of the adopted magnetic braking.
The second scenario corresponds to the case of short initial orbital periods ($\sim$3~days), and is characterized by a decrease in the orbital period at the end of the evolution of the sequences that consider the CARB formula.

Also, for intermediate and long initial orbital periods, the binary components do not come closer before RLOF begins if the \citetalias{Rappaport1983MB} magnetic braking is considered.
For the case in which the CARB magnetic braking is considered, the components get closer before the beginning of the RLOF, and the difference in relation to the prescription of R83 is as large as the shorter the $P_\text{i}$.
Finally, in the case of short initial orbital periods, both prescriptions for magnetic braking (\citetalias{Rappaport1983MB} and CARB) considerably decrease the separation of the binary components before the RLOF begins.
In summary, the effect of magnetic braking increases with the decreasing of orbital periods for both prescriptions, but it alters differently along initial orbital periods for each prescription.

For $P_\text{i}=100~\text{d}$, both \citetalias{Rappaport1983MB} and CARB prescriptions results in diverging systems; and for $P_\text{i}=3~\text{d}$, they both converge (see Fig.~\ref{fig:plot-Md-P} and Section~\ref{subsec:PeriodBif}).
The relation between the initial and final orbital periods for each magnetic braking prescription is, however, completely different.
Looking for the limit that separates the converging from the diverging systems (i.e, $P_\text{i}=P_\text{f}$), we find $\sim 3$~d for the \citetalias{Rappaport1983MB} magnetic braking but 20~d if the CARB prescription is used.
With the only exception of the $P_\text{i}=100~\text{d}$ models, in all other cases the use of the CARB prescription causes the mass transfer to start earlier when compared to the \citetalias{Rappaport1983MB} prescription (second panel in Fig.~\ref{fig:plot-age-all5}).
Also, the duration of the mass transfer is longer for shorter $P_\text{i}$.
In fact, for $P_\text{i}=100~\text{d}$ the mass transfer phase is so fast that we can barely distinguish it in the figure.

We now examine how the variables that determine the intensity of magnetic braking evolve.
During most of the evolution, the \citetalias{Rappaport1983MB} prescription results in stronger magnetic braking than the CARB prescription (first panel in Fig.~\ref{fig:plot-age-all5}).
The exception occurs during mass transfer, where the CARB braking prescription becomes more intense.
This inversion can occur up to about 500~Myr before the mass transfer begins.
Although these two prescriptions for magnetic braking differ for systems of any initial orbital period, their effects are much more intense in systems of short and intermediate orbital periods ($P_\text{i} \sim 3\text{--}50~\text{d}$) than in long orbital periods ($P_\text{i} \sim 100~\text{d}$).

The $P_\text{i}=100~\text{d}$ sequences show minimal differences in the evolution of the orbital period and donor mass (Fig.~\ref{fig:plot-Md-P}).
This occurs because the rotation rate, the radius of the donor, and the the donor mass evolve in a very similar way in both prescriptions, for this initial orbital period (third and fifth panels in Fig.~\ref{fig:plot-age-all5}; the evolution of the mass of the donor is not show, for simplicity).
The small difference in the orbital period behaviour during the evolution of these sequences is due to the role of the size of the convective zone in the CARB prescription, which, in turn, affects the magnetic braking (first panel in Fig.~\ref{fig:plot-age-all5}).
For the $P_\text{i}=100~\text{d}$ sequence with the CARB formula, the convective zone at the beginning of the RLOF accounts for $\sim 0.82~M_{\odot}$ of the model.
Finally, the angular momentum loss from magnetic braking has a very limited impact on the evolution of binaries with long orbital periods.
During the pre-RLOF evolution of the $P_\text{i}=100~\text{d}$ systems, magnetic braking has an almost null contribution in both prescriptions.
On the other hand, for $P_\text{i}=3~\text{d}$, magnetic braking is the dominant mechanism in the \citetalias{Rappaport1983MB} prescription; and has an increasing contribution in the CARB prescription, dominating from 2~Gyr before RLOF onwards.

In the second scenario, $P_\text{i}=3~\text{d}$,
the mass transfer begins when the donor star has just left the main sequence and
the convective zone is not extended enough ($\sim 0.25~M_{\odot}$ in the CARB sequence) to contribute significantly with magnetic braking.
Thus, the consideration of the size of the convective zone foreseen by the CARB prescription has little effect.
Until the beginning of the RLOF, the wind mass loss and the radius of the donor star behave similarly in both prescriptions, assuming values of $10^{-11.5}~M_{\odot}\, \text{yr}^{-1}$ and $2~R_{\odot}$ immediately before the beginning of the mass transfer, respectively (fourth and fifth panels in Fig.~\ref{fig:plot-age-all5}). 
On the other hand, the rotation rate in the
CARB \citepalias{Rappaport1983MB} prescription is 32~km/s (59~km/s)
at age 5.2~Gyr (third panel in Fig.~\ref{fig:plot-age-all5}).
Since the donor mass and the donor radius behave similarly in this case, contributions are limited to the rotation rate, the wind mass loss, and the convection turnover time in this case.
These are the contributions that make the mass transfer start when the system has an orbital period of 1.1 days (CARB) and 0.9 days (\citetalias{Rappaport1983MB}).
When the \citetalias{Rappaport1983MB} prescription is used in the $P_\text{i}=3~\text{d}$ system, the rotation rate, the radius of the donor star, and the donor mass react smoothly to the mass loss (third and fifth panels in Fig.~\ref{fig:plot-age-all5}).
Thus, the magnetic braking in this case also remains approximately constant $\log_{10}(|\dot{J}_\text{mb}|)=35$ during a few Gyr after the mass transfer (first panel in Fig.~\ref{fig:plot-age-all5}).
On the other hand, when the CARB prescription is considered, the donor star contracts, and the wind mass loss decays after the end of the RLOF, causing the magnetic braking to be reduced to $\log_{10}(|\dot{J}_\text{mb}|)=28$.
In the case where $P_\text{i}=3~\text{d}$, although the \citetalias{Rappaport1983MB} prescription results in a considerably longer RLOF, the CARB prescription reaches
$\dot{M}=10^{-7}~M_{\odot}~\text{yr}^{-1}$,
while the previous one only
$\dot{M}=10^{-8.2}~M_{\odot}~\text{yr}^{-1}$.
In addition, the moment the mass transfer ends, the system using the \citetalias{Rappaport1983MB} prescription has an increasing orbital period, while the CARB prescription has a decreasing orbital period.
At this point, the contributions of magnetic braking and mass loss to the total angular momentum loss are around 88 (98) and 12 (2) per cent for \citetalias{Rappaport1983MB} (CARB), respectively.
Gravitational radiation will dominate the angular momentum loss only about 2~Gyr after the RLOF terminates.
Furthermore, at this point, the donor star has a radius about 2.3 times larger for \citetalias{Rappaport1983MB} than for CARB.
In both cases the donor radius remains close to the Roche lobe after the end of the mass transfer, but for \citetalias{Rappaport1983MB} these two quantities are increasing, and for CARB they are decreasing.
The analysis of these factors makes it clear that the evolution of the donor star and the binary system combine differently for each prescription of magnetic braking, which will be presented in more detail in the next section.

Although the $P_\text{i}=20~\text{d}$ sequences behave similarly to the $P_\text{i}=100~\text{d}$ sequences, they differ in the evolution of the orbital period before the RLOF.
For $P_\text{i}=20~\text{d}$, the RLOF starts when $P=18~\text{d}$ in the case of \citetalias{Rappaport1983MB} magnetic braking and when $P=5.7~\text{d}$ in the case of the CARB prescription (see Fig.~\ref{fig:plot-Md-P}).
This means that the entire mass transfer takes place with the stars much closer together when CARB magnetic braking is considered.
This reinforces the fact that each magnetic braking prescription leads to a different shrinkage of the orbit, and therefore to a different evolutionary stage of the donor and orbital separation at the onset of the mass transfer.
Both the radius of the donor star and the loss of mass by winds reach higher values (around $60~R_{\odot}$ and $10^{-8}~M_{\odot}\, \text{yr}^{-1}$, respectively) during mass transfer when \citetalias{Rappaport1983MB} braking is considered.
On the other hand, when CARB braking is considered, the radius of the donor star remains stable around $9~R_{\odot}$ during mass transfer.
In addition, the loss of mass by winds intensifies in this case and remains around $10^{-9.5}~M_{\odot}\, \text{yr}^{-1}$ for about 1.5~Gyr.
For both braking prescriptions, these two variables drop dramatically as soon as the RLOF ends.
The consequence is that the mass transfer lasts about three times longer when the CARB prescription is considered.

For any $P_\text{i}$, the CARB magnetic braking is more intense than the \citetalias{Rappaport1983MB} prescription at a time when the mass transfer rate is maximum (first panel in Fig.~\ref{fig:plot-age-all5}).
In addition, for any initial orbital period, the CARB magnetic braking is less intense than the \citetalias{Rappaport1983MB} prescription braking after mass transfer.

\subsection{Orbital period evolution and period bifurcation}
\label{subsec:PeriodBif}

There is a critical initial orbital period --- called the bifurcation period --- that separates the systems in
converging ($P_\text{f} < P_\text{i}$)
and
diverging ($P_\text{f} > P_\text{i}$).
The converging systems are the ones that, after the Roche lobe overflow, evolve with decreasing orbital period until the donor star becomes degenerate and an ultra-compact binary is formed.
The diverging systems are the ones that, after the RLOF, evolve with increasing orbital period and a wide detached binary is formed
\citep{PylyserSavonije1988,PylyserSavonije1989}.
In theoretical models,
the bifurcation period depends on the strength of the magnetic braking
\citep[e.g.,][]{TaurisHeuvel2006}.
A systematic study made by \citet{MaLi2009}
considering systems with a $1.4~M_{\odot}$ neutron star and a $0.5\text{--}2~M_{\odot}$ donor star
found that the strength of the magnetic braking is the dominant factor in determining the value of bifurcation period compared with mass loss.

\begin{figure}
	\includegraphics[width=\columnwidth]{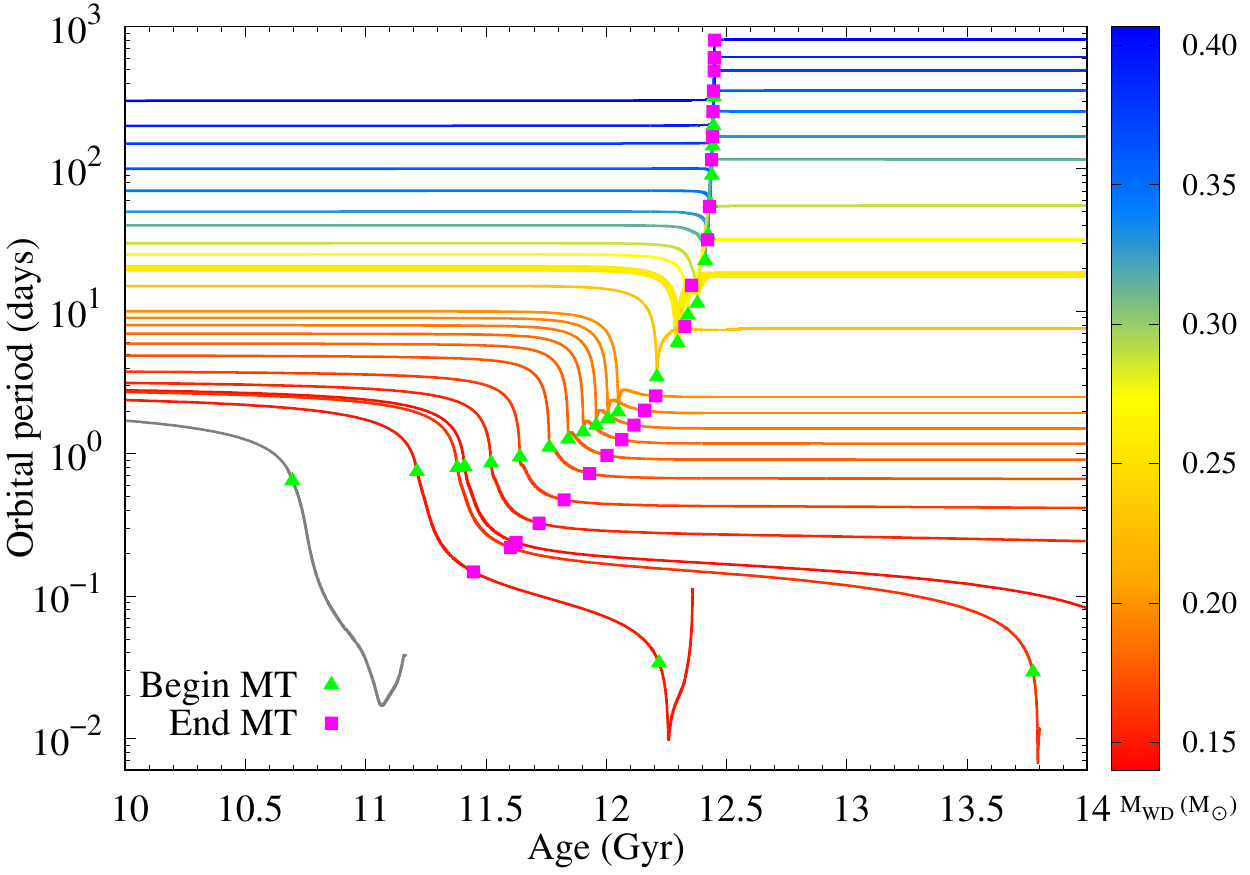}
    \caption{The evolution of the orbital period as a function of age for selected models between $2.7 \leqslant P_\text{i}/\text{d} \leqslant 300$. Initial orbital periods, from top to bottom: 300, 200, 150, 100, 70, 50, 40, 30, 25, 20, 15, 10, 9, 8, 7, 6, 5, 4, 3.5, 3.25, 3.2, 3 and 2.7 days. The bifurcation period occurs between 20 and 25~d. The extra-thick line marks the first convergent system. Above it, all systems are divergent. The beginning and end of the mass transfer are indicated by green triangles and pink squares, respectively. The colour of each line indicates the mass of the donor at the end of the mass transfer. The sequence shown in grey never becomes detached. The initial configuration is $M_{\text{d,i}} = 1.0~M_{\odot}$ and $M_{\text{a,i}} = 1.4~M_{\odot}$ for all sequences. The enhanced CARB magnetic braking is considered. Convergent binaries will continue to contract their orbits, forming a cataclysmic variable or an ultra-compact X-ray binary system. Divergent binaries will become relatively wide systems containing a recycled NS and a He or CO WD.}
    \label{fig:plot-Pbif}
\end{figure}

In this section we expand our model grid using the CARB prescription of the magnetic braking into the initial orbital period parameter space.
Fig.~\ref{fig:plot-Pbif}
shows the evolution of the orbital period as a function of age for the CARB prescription,
for initial masses $M_{\text{d,i}} = 1.0~M_{\odot}$ and $M_{\text{a,i}} = 1.4~M_{\odot}$.
The initial orbital periods range from 2.7 to 300 days.
Because of the initial donor mass and metallicity, no model shows a significant orbital period variation before a model age of 10~Gyr.
The shorter the $P_\text{i}$, the sooner the systems starts mass transfer.
Up to this point we are only discussing models without element diffusion.
However, we draw attention to the fact that this does not affect the main results discussed so far, as diffusion has a small effect on the quantities studied.
The effects of diffusion will be discussed from Sec.~\ref{subsec:ELMmassPi} on-wards.

Using the \citet[][\citetalias{Rappaport1983MB}]{Rappaport1983MB} magnetic braking,
\citet{Istrate2016Models-ads} found that the bifurcation period that separates the converging systems from the diverging ones occurs between 2.75 and 2.8 days if
$M_{\text{d,i}} = 1.0~M_{\odot}$
and
$M_{\text{a,i}} = 1.4~M_{\odot}$.
Considering the CARB magnetic braking prescription of
\citetalias{Van2019},
a $P_{\text{i}} = 20~{\text{d}}$ system
with the same initial masses
is still a convergent system.
This corresponds to a $0.255~M_{\odot}$ ELM white dwarf --- the thickest line in Fig.~\ref{fig:plot-Pbif}.

We should emphasise that even in the lower limit of orbital period, our models
(as shown in Appendix~\ref{sec:Appendix-CompMod}) are completely detached at a model age of 14~Gyr.
Simulations considering shorter initial orbital periods
still have a small rate of mass transfer
($\dot{M}>10^{-10}~M_{\odot}~\text{yr}^{-1}$) at the final computed age and therefore are not part of our model grid.
In such cases, the donor star is expected to be completely consumed, i.e., transfer all its mass to the neutron star within a few billion years --- or they may merge due to emission of gravitational radiation, similar to the known ultra-compact binaries (bottom three sequences in Fig.~\ref{fig:plot-Pbif}).

The bifurcation period is shifted to longer ones (from 2.75--2.8 to 20--25~days) when the CARB magnetic braking is considered.
I.e., the CARB magnetic braking allows us to get ELM WDs models with masses as low as $0.26~M_{\odot}$ in converging binary systems even with initial orbital periods as long as 20 days,
which is not possible with the magnetic braking of \citetalias{Rappaport1983MB}.
This is important because it shows that the entire extension of low-mass and ELM WDs in systems with pulsars can be obtained via RLOF evolution from a more uniform distribution of initial orbital periods (on a logarithmic scale), without favouring only the most massive ones.

Although not the focus of this work, the upper and lower sequences in Fig.~\ref{fig:plot-Pbif} show that the use of CARB magnetic braking makes it possible to form UCXB systems and wide-orbit binary millisecond pulsars, which is not possible with the \citetalias{Van2018} prescription.

At this point, we should consider whether there are new problems appearing with the use of CARB, since even the binaries with $P_\text{i}=20$~d can produce ELM WDs in millisecond pulsar systems.
In fact, the analysis of \citet{Istrate2014} indicates with a high level of confidence that the distribution of orbital periods of observed recycled pulsars with He WD companions in the Galactic field is not compatible with the simulations that use the R83 prescription.
They pointed out that the range of initial orbital periods that lead to the formation of this type of system must be expanded.
In addition, the $P_\text{i}$--$M_\text{d,f}$ relation we found is much closer to the expected log-normal orbital period distribution \citep[see e.g.,][]{Duchene2013-ads,Tutukov2020}
than when using the \citetalias{Rappaport1983MB} prescription.
Thus, the results we found using the CARB prescription are encouraging and a study comparing these results with simulations of binary population synthesis looks promising.

\subsection{ELM with neutron stars}
\label{subsec:ELM+NS}

In this subsection we expand our study of systems with point mass accretors of $1.4~M_\odot$.
\subsubsection{Initial orbital period and final mass}
\label{subsec:ELMmassPi}

%
\begin{figure}
	\includegraphics[width=\columnwidth]{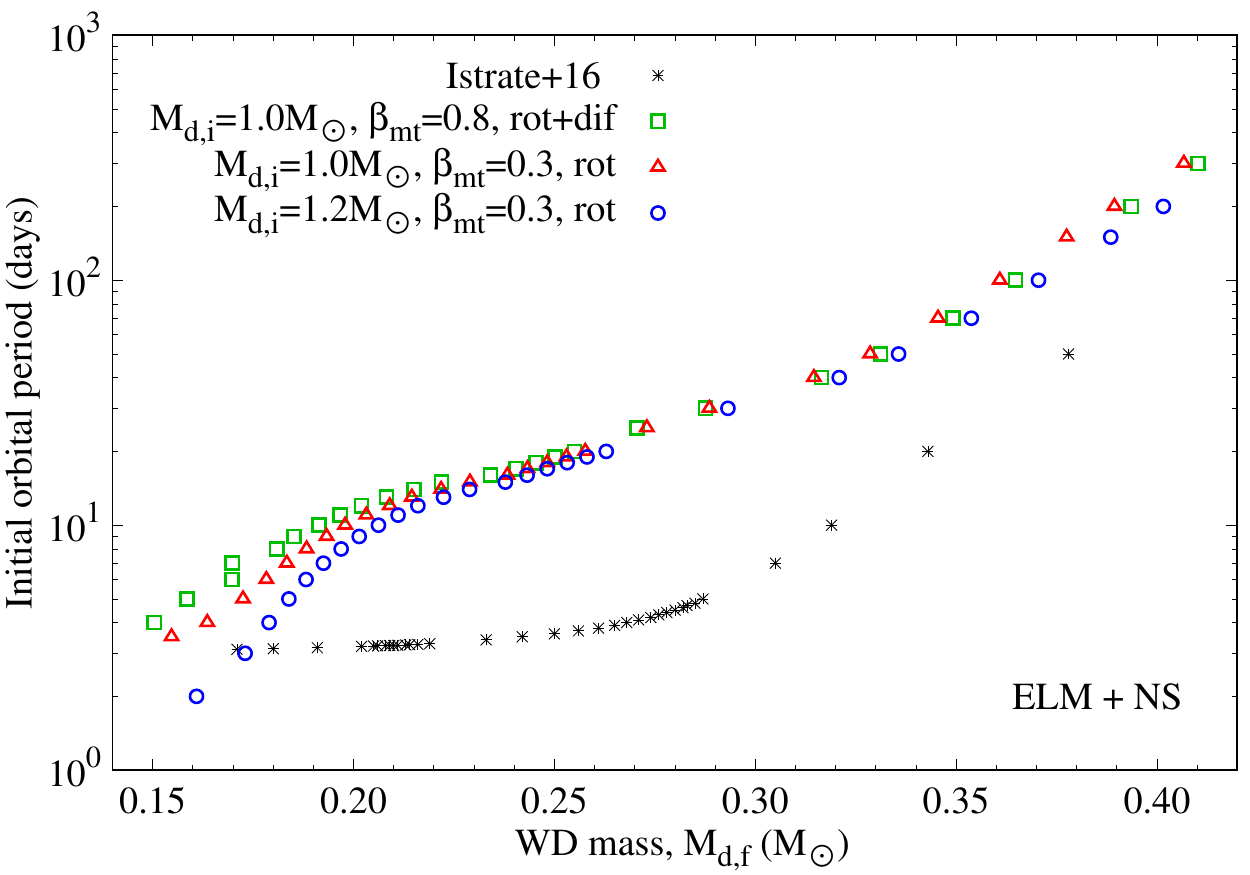}
    \caption{The relation between the initial orbital period and the ELM WD mass at the end of 14~Gyr evolution for different setups. \citet{Istrate2016Models-ads} LMXB models using the \citet{Rappaport1983MB} magnetic braking prescription are shown in black ``$\ast$'' signs.
    Green ``$\square$'' signs: $M_\text{d,i}=1.0~M_{\odot}$, 20 per cent accretion efficiency, rotation plus diffusion.
    Red ``$\triangle$'' signs: $M_\text{d,i}=1.0~M_{\odot}$, 70 per cent accretion efficiency, rotation only.
    Blue ``$\circ$'' signs: $M_\text{d,i}=1.2~M_{\odot}$, 20 per cent accretion efficiency, rotation only.
    All configurations have $M_\text{a,i}=1.4~M_{\odot}$.}
    \label{fig:plot-M1Porb-mass}
\end{figure}

In this section we show how the relation between the initial orbital period and the final white dwarf mass is modified when we change the initial mass of the donor star and the neutron star mass accretion efficiency.
In Fig.~\ref{fig:plot-M1Porb-mass} we depict the relation between the initial orbital period ($y$-axis) and the low-mass/ELM final mass ($x$-axis).
Red triangles correspond to models with $M_\text{d,i}=1.0~M_{\odot}$, 70 per cent accretion efficiency (i.e., $\beta_\text{mt}=0.3$), and that take into account rotation only.
Blue circles are for $M_\text{d,i}=1.2~M_{\odot}$; and green squares are for 20 per cent accretion efficiency (i.e., $\beta_\text{mt}=0.8$) with both rotation and diffusion.
For comparison, \citet{Istrate2016Models-ads} LMXB models using the \citet{Rappaport1983MB} magnetic braking prescription are shown in black ``$\ast$'' signs.

As we already mentioned in Sec.~\ref{sec:Intro}, the empirical treatment of the magnetic braking by \citet{Rappaport1983MB} leads to a fine-tuning of the order of a dozen minutes in the initial orbital period to reproduce the observed orbital periods of millisecond pulsars in compact ($2 < P / \text{h} < 9$) binaries with He WD companions of mass $\lesssim 0.20~M_{\odot}$ \citep{Istrate2014}.

It is notable that each prescription for magnetic braking has a completely different pattern in the $P_\text{i}$--$M_\text{d,f}$ plane.
For final donor masses between 0.17 and $0.25~M_{\odot}$, we can see in Fig.~\ref{fig:plot-M1Porb-mass} that the range of corresponding initial orbital periods is extremely narrow (between 2 and 4 days) for the prescription of \citetalias{Rappaport1983MB}.
On the other hand, when the CARB \citepalias{Van2019} formulation is considered, the same range of final masses is obtained for initial orbital periods between 3 and 20 days.
Thus, the use of the CARB prescription does not require a fine-tuning of initial periods for the formation of ELM white dwarfs.
Using initial masses $M_{\text{d,i}} = 1.0~M_{\odot}$ and $M_{\text{a,i}} = 1.4~M_{\odot}$,
we were able to produce detached white dwarf systems within the range
$3.25 \leqslant P_\text{i}/\text{d} \leqslant 300$ (see Fig.~\ref{fig:plot-PiPf}),
which corresponds to ELM and low-mass WDs with masses in the range
$0.1456 \leqslant M_{\text{d,f}}/M_{\odot} \leqslant 0.4067$.

We will now consider only the sequences that use CARB \citepalias{Van2019} magnetic braking.
In Fig.~\ref{fig:plot-M1Porb-mass}, the setup discussed in the previous subsection ($M_\text{d,i}=1.0~M_{\odot}$, $\beta_\text{mt}=0.3$) is shown in red triangles.
The first comparison concerns an initially more massive donor, with $1.2~M_{\odot}$ (blue circles in Fig.~\ref{fig:plot-M1Porb-mass}).
There is a systematic shift of the final mass towards larger masses, for the same initial period $P_\text{i}$, in comparison to the case when the donor mass is $M_\text{d,i}=1.0~M_{\odot}$.
The difference in final masses increases for shorter initial orbital periods.
For example, we find the final donor mass to be $0.1790~M_{\odot}$ if $M_\text{d,i}=1.2~M_{\odot}$, but $0.1636~M_{\odot}$ if $M_\text{d,i}=1.0~M_{\odot}$.
For $P_\text{i} \leqslant 11$~d, no sequence undergoes hydrogen shell flashes, regardless of the initial mass of the donor.
The configuration with $M_\text{d,i}=1.2~M_{\odot}$ is the only one where it is possible to obtain detached systems for $P_\text{i} < 3$~d.

Using a $M_\text{d,i}=1.2~M_{\odot}$ donor instead of $1.0~M_{\odot}$ does not significantly affect the binary evolution.
The difference in the ELMs final mass is due to the more massive model being able to burn more H into He before mass transfer begins.
For example, for the $P_\text{i}=20~\text{d}$ case, this is reflected in the He core to be $0.0054~M_{\odot}$ more massive for the initially more massive donor.
This difference increases to $0.015~M_{\odot}$ for $P_\text{i}=300~\text{d}$.
Note that for this metallicity ($Z=0.02$) and $M_\text{d,i}=1.0~M_{\odot}$, because of the long main sequence lifetime, it is difficult to produce (pre-)ELMs in less than 10--11 Gyr.
Thus, younger (pre-)ELMs require either lower metallicity or initially more massive donors.

The next comparison concerns the accretion efficiency to the neutron star and the use of diffusion (green squares in Fig.~\ref{fig:plot-M1Porb-mass}).
The effect of diffusion on the relation shown in Fig.~\ref{fig:plot-M1Porb-mass} is negligible for $M_\text{d,f} \gtrsim 0.3~M_{\odot}$.
For $M_\text{d,f} \lesssim 0.3~M_{\odot}$, the diffusion of elements leads to slightly smaller final masses, given a $P_\text{i}$, because diffusion brings more fuel to the burning zone, in addition to leading to more hydrogen shell flashes.

On the other hand, different accretion efficiencies have a significant effect only for $M_\text{d,f} \gtrsim 0.3~M_{\odot}$.
In this case, low accretion efficiency results in low-mass/ELM WDs with a slightly higher final mass, for a given $P_\text{i}$.
This is because the mass transfer rate reaches slightly higher values as the accretion efficiency to the point-like neutron star is higher.
Another consequence is that, for a given $P_\text{i}$, the orbital separation after the mass transfer is shorter when the accretion efficiency is higher.

The occurrence of different numbers of hydrogen shell flashes can also be noted in Fig.~\ref{fig:plot-M1Porb-mass} when looking at the systems with diffusion (green squares).
In this case, the $P_\text{i}=7$ and 6~days systems have practically the same donor final mass ($0.1698~M_{\odot}$ and $0.1697~M_{\odot}$, respectively), but the $P_\text{i}=7$~d sequence undergoes one hydrogen shell flash more than the other.
This causes a discontinuity in the form of a step in the curve formed by the models in this figure.
In summary, the variation of the initial mass of the donor stars and efficiency rate to the accreting star, and the consideration of diffusion of elements cause significant effects in the $P_\text{i}$--$M_\text{d,f}$ plane for $P_\text{i} \lesssim 10~\text{d}$ or, equivalently, $M_\text{d,f} \lesssim 0.2~M_{\odot}$.

\subsubsection{Initial and final orbital period}
\label{subsec:PiPf}

\begin{figure}
	\includegraphics[width=\columnwidth]{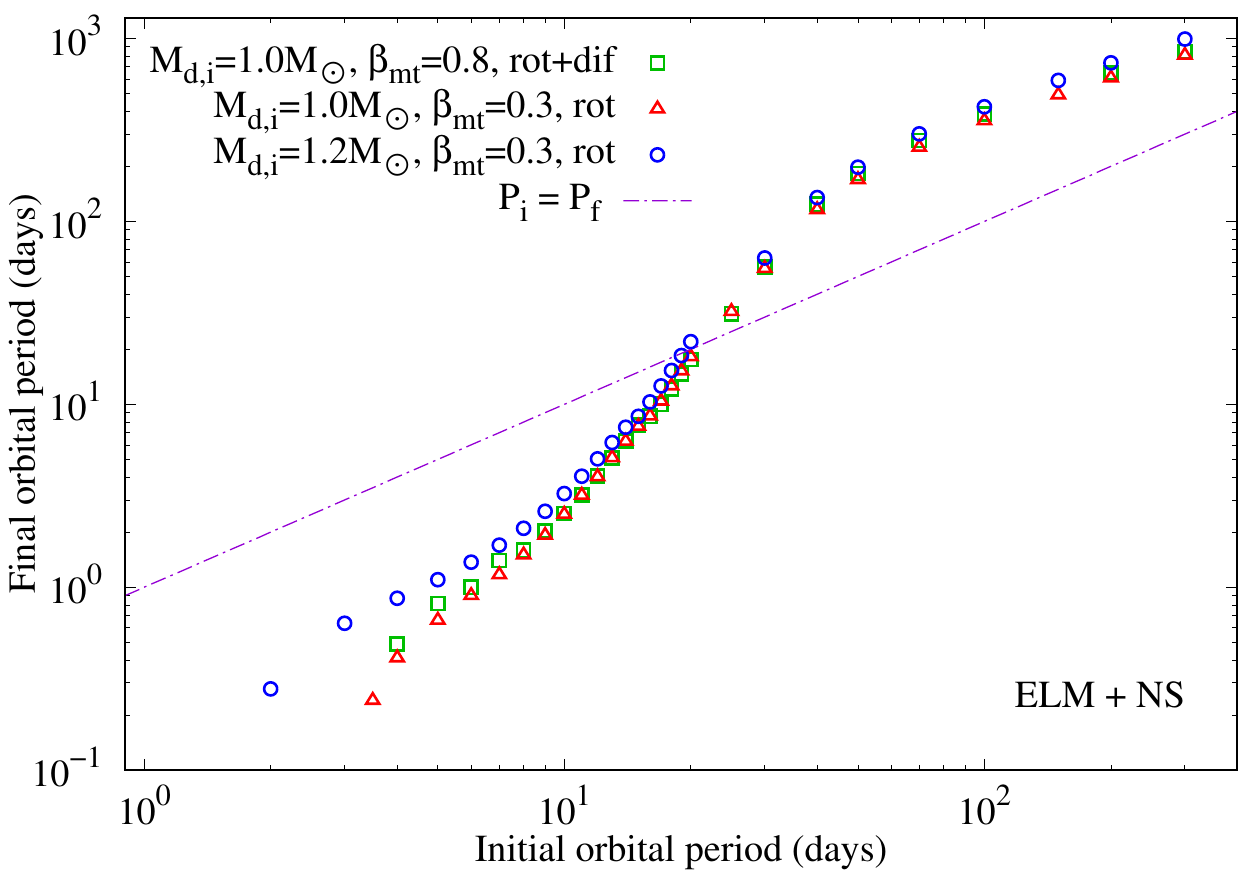}
    \caption{The relation between the final ($P_\text{f}$) and the initial orbital period ($P_\text{i}$). Each initial configuration is as follows.
    Green ``$\square$'' signs: initial donor mass $M_\text{d,i}=1.0~M_{\odot}$, 70 per cent accretion efficiency, rotation plus diffusion.
    Red ``$\triangle$'' signs: $M_\text{d,i}=1.0~M_{\odot}$, 20 per cent accretion efficiency, rotation only.
    Blue ``$\circ$'' signs: $M_\text{d,i}=1.2~M_{\odot}$, 20 per cent accretion efficiency, rotation only.
    All configurations have $M_\text{a,i}=1.4~M_{\odot}$.
    The purple dot-dashed line serves as an indicator to distinguish between convergent and divergent systems.}
    \label{fig:plot-PiPf}
\end{figure}

In this section we show how the initial--final orbital period relation is modified when we change the initial mass of the donor star and the mass accretion efficiency to the neutron star.
In Fig.~\ref{fig:plot-PiPf} we show the final orbital period ($y$-axis) as a function of the initial orbital period ($x$-axis).
Red triangles are for sequences with $M_\text{d,i}=1.0~M_{\odot}$, 70 per cent accretion efficiency (i.e., $\beta_\text{mt}=0.3$), and that takes into account rotation only.
Similarly, blue circles are for $M_\text{d,i}=1.2~M_{\odot}$; and green squares are for 20 per cent accretion efficiency (i.e., $\beta_\text{mt}=0.8$) with both rotation and diffusion.
The purple dot-dashed line serves as an indicator to distinguish between convergent and divergent systems.

Fig.~\ref{fig:plot-PiPf} shows that the division between convergent and divergent systems is around $P_\text{i} = 20~\text{d}$, regardless of the initial configuration.
Therefore, the pattern of orbital evolution shown in Fig.~\ref{fig:plot-Pbif} for that specific initial configuration is similar also for the other configurations presented here.

For any initial orbital period, an initially more massive donor always leads to a wider binary, regardless of the accretion efficiency to the neutron star.
In fact, it is possible to identify a tendency for initially short ($P_\text{i} < 10~\text{d}$) and long ($P_\text{i} > 40~\text{d}$) orbital period systems.
In these cases, we find that, for a given initial orbital period, higher accretion efficiency leads to shorter final orbital periods; and an initially more massive donor leads to longer final orbital periods.
For orbital periods closer to the bifurcation period, however, there is no clear trend, and the $P_\text{i}$--$P_\text{f}$ relations are quite similar for each configuration studied.

These results can be understood as the mass transfer begins earlier for systems that have a donor with a 1.2~$M_{\odot}$ (5--6~Gyr) compared to a 1.0~$M_{\odot}$ donor (11--12~Gyr) due to their main sequence lifetimes. Angular momentum loss mechanisms have more time to act on systems with less massive donors, leading to mass transfer beginning when the components of the binary system are closer.

In Fig.~\ref{fig:plot-PiPf}
we compare models with diffusion and low accretion efficiency (green squares) against models without diffusion and high accretion efficiency (red triangles).
We found that diffusion of elements and accretion rate efficiency affect evolution at different times.
Until the beginning of the mass transfer, neither the inclusion of diffusion or different accretion rates affect significantly the evolution.
At the end of mass transfer, low accretion efficiency to the neutron star makes sequences with $\beta_\text{mt} = 0.8$ with their components closer together.
Still, for a given $P_\text{i}$, the masses of each configuration are similar at that time.
What happens next for sequences of $M_\text{d,f} \lesssim 0.3~M_{\odot}$ depends on the inclusion or not of the diffusion of elements.
For most non-diffusing sequences, none or one hydrogen shell flashes occurs.
On the contrary, in most sequences with diffusion, there are two or three hydrogen shell flashes.
As the loss of angular momentum during hydrogen shell flashes is dominated by mass loss for a more massive accretor, each hydrogen shell flash increases orbital separation.
Thus, the inclusion of diffusion of elements tends to decrease the final donor mass and increase the orbital separation of the components.

We also found a clear relation between the final orbital period and the rotation rate of the white dwarf.
For models with diffusion, systems with a short initial orbital period ($P_\text{i} \lesssim 15~\text{d}$) present greater synchronization with the orbit, at the end of the evolution.
In such cases, the ratio between the rotation rate of the ELM white dwarf and the orbital period of the system assumes values between 1 and 0.1.
On the other hand, this ratio in $P_\text{i} \gtrsim 20~\text{d}$ systems is
$P_\text{rot}/P_\text{f}\simeq 0.1$--$10^{-3}$, indicating white dwarfs rotating more slowly than the orbital period.
Looking at the $P_\text{i}$--$P_\text{f}$ diagram in Fig.~\ref{fig:plot-PiPf}, we notice that this value of the initial orbital period (20~d) coincides with the bifurcation period.
Thus, convergent systems are more likely to have synchronization between the rotation of the white dwarf and the orbit.
This occurrence might be a tool to observationally estimate the convergence period.
Even for systems with a shorter initial orbital period, we find that from $\sim$1~Gyr after the end of RLOF onwards, the time needed for synchronization exceeds the age of the Universe. This means that we should not expect tidal forces to change the rotation of the newly formed low-mass/ELM WDs.

Note that the bottom-left corner of Fig.~\ref{fig:plot-PiPf} is dominated by systems with $P<1~\text{d}$.
Such close systems are strong candidates to be observable in gravitational waves \citep[e.g.,][]{ChenTauris2021}.
Simple estimates of the amplitude, the characteristic strain, and the frequency of the gravitational waves emitted by our models at final age are described in Appendix~\ref{sec:Appendix-GW} and presented in Appendix~\ref{sec:Appendix-CompMod}.

\subsubsection{Final mass--orbital period relation}
\label{subsec:MfPf}

The determination of relations between the orbital period and the mass of low-mass and ELM WDs in systems with neutron stars is of great interest because this relation can be used to examine the evolutionary channel for such a binary.
These relations allow the estimation of the masses of ELM WDs from the orbital period of the binary system, which is, in general, easier to measure and independent of the $T_\text{eff}$ and $\log (g)$ determinations.

The cores masses and radii of low-mass stars in the red giant branch follow a tight, well-known relation \citep{Refsdal1971,Webbink1983,Joss1987}. For a red giant donor in a binary system, its radius is approximately equal to its Roche lobe radius during the mass transfer phase. The latter, in turn, depends on the binary separation and on the mass ratio. At the end of the mass transfer phase, the H-rich envelope is almost completely removed, and the final mass of the donor star is approximately the mass of its He core. Therefore, the final mass of a degenerate-core donor and the orbital period are correlated quantities.

\begin{figure*}
\begin{minipage}{\textwidth}
	\includegraphics[width=\columnwidth]{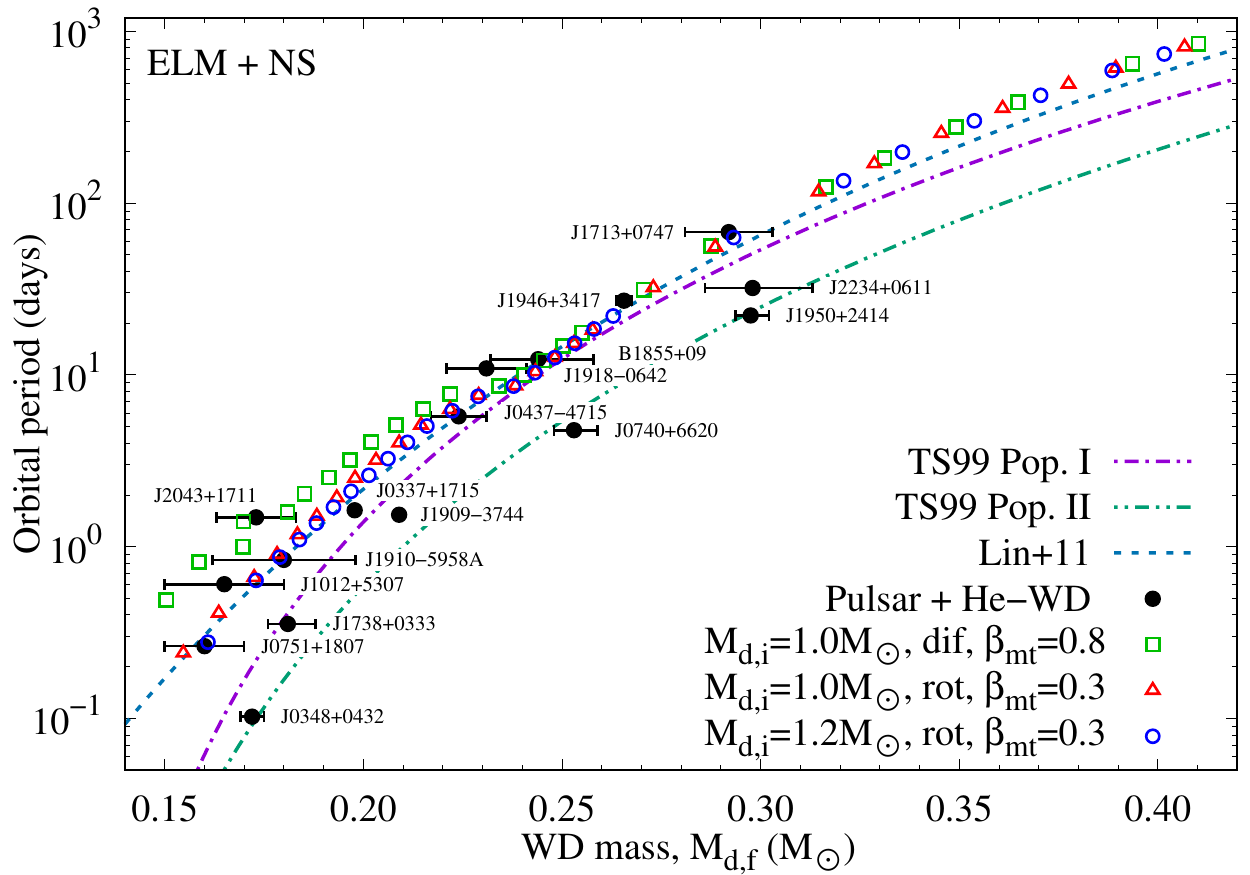}
    \caption{The relation between the low-mass/ELM WD final mass and the orbital period.
    Each initial configuration is as follows.
    Green ``$\square$'' signs: $M_\text{d,i}=1.0~M_{\odot}$, 70 per cent accretion efficiency, rotation plus diffusion.
    Red ``$\triangle$'' signs: $M_\text{d,i}=1.0~M_{\odot}$, 20 per cent accretion efficiency, rotation only.
    Blue ``$\circ$'' signs: $M_\text{d,i}=1.2~M_{\odot}$, 20 per cent accretion efficiency, rotation only.
    All configurations have $M_\text{a,i}=1.4~M_{\odot}$.
    The dash-dotted curves represents the theoretical relations from \citet[][TS99]{Tauris1999} and the dotted curve from \citet[][Lin+11]{Lin2011}.
    Observational data (black dots) are from pulsar + He-WD systems, with 16 systems in total.
    Dots without an uncertainty bar mean that the uncertainty is less than the dot size.
    }
    \label{fig:plot-MfPf}
\end{minipage}
\end{figure*}

Fig.~\ref{fig:plot-MfPf} shows the final mass of donor vs. period ($M_\text{d,f}$--$P_\text{f}$) relation for all computed models with a $1.4~M_\odot$ point mass accretor.
Although the CARB \citepalias{Van2019} prescription for the magnetic braking completely changes the relation between $P_\text{i}$ and $M_\text{d,f}$, the relation between the final period ($P_\text{f}$) and final donor mass ($M_\text{d,f}$) is much less affected in relation to the models calculated with the \citet[][\citetalias{Rappaport1983MB}]{Rappaport1983MB} magnetic braking formalism
\citep{Istrate2016Models-ads}.
The use of the CARB prescription produces models that maintain agreement with other theoretical adjustments and also with observational data.

Our results are in good agreement with the \citet{Lin2011} theoretical fit, which is based on low and intermediate-mass X-ray binaries models computed with gravitational radiation, mass loss, and \citetalias{Rappaport1983MB} magnetic braking contributions to the angular momentum loss.
In particular, our models for $M_\text{d,f} \lesssim 0.30~M_{\odot}$ and $\beta_\text{mt} = 0.3$ are in excellent agreement with the \citet{Lin2011} fit.
For $\beta_\text{mt} = 0.8$, although the agreement with the fit is good for $M_\text{d,f} \gtrsim 0.23~M_{\odot}$, we found orbital periods slightly longer than \citet{Lin2011} if $M_\text{d,f} \lesssim 0.23~M_{\odot}$, for a given mass.
Between $M_\text{d,f} = 0.23$ and $0.26~M_{\odot}$, our models are in good agreement with both \citet{Lin2011} and \citet[][Pop. I]{Tauris1999} fits.
It is important to note that these fits were made based on a broader initial donor masses distribution than we are considering here.

Although the angular momentum losses of the systems are calculated differently in each work, the $M_\text{d,f}$--$P_\text{f}$ relation is determined by the state of the He cores, therefore it should not be changed with the angular momentum loss.
In fact, \citet{ChenHanDeca2013} have shown that the mass transfer efficiency and the way that angular momentum is lost impose only a small influence on the $M_\text{d,f}$--$P_\text{f}$ relations.
On the other hand, the mixing length and the metallicity affect this relation \citep[e.g.,][]{Rappaport1995,Tauris1999}.
Also, the calibration of other internal processes --- such as the opacity and the convective overshooting --- can influence the radius of red giant stars and hence the $M_\text{d,f}$--$P_\text{f}$ relation.
As already mentioned, we found that the inclusion of diffusion of elements leads to less massive donors, given an initial orbital period.
Analysing the effects of including diffusion and low accretion efficiency separately, we find that both decrease the final mass of the donor, for a given initial orbital period, although the former has a more pronounced effect than the latter.
Thus, in Fig.~\ref{fig:plot-MfPf} we attribute the discrepancy between the green squares and the fit of \citet{Lin2011} primarily to the effects of diffusion.

An important result is that the relation between the orbital period and the final mass of ELM WDs is not significantly affected when we consider different initial donor masses (1.0 and $1.2~M_{\odot}$), although considerable differences are expected for donors of initial masses 1.3--$1.6~M_{\odot}$ \citep{Istrate2014}.
On the other hand, for $M_\text{d,f} \lesssim 0.23~M_{\odot}$, low accretion efficiencies always lead to more compact systems, for a given mass.

Fig.~\ref{fig:plot-MfPf} also shows observational data, which include sixteen systems of (millisecond) pulsar plus He-WDs with individual component mass measurements\footnote{Observational data of pulsars were taken from \url{https://www3.mpifr-bonn.mpg.de/staff/pfreire/NS_masses.html}
\citep[see also][]{Ozel2016,Antoniadis2016},
\citet[][PSR J1946+3417]{Barr2017}, and
\citet[][PSR J0740+6620]{Fonseca2021}.}.
Each pulsar companion is shown with a black dot and an $\pm 1\sigma$ uncertainty bar in the mass measurement.
Dots without an uncertainty bar have uncertainty in the measure of mass smaller than the dots size.
This list contains only systems in which there is no mass transfer nor significant mass loss observed.

Instead of commenting on each of the pulsars separately, we highlight some systems that deviate most from the theoretical estimates of \citet{Tauris1999,Lin2011}.
It is well known that, for a given final mass, lower metallicities lead to shorter final orbital periods (TS99 Pop. II curve in Fig.~\ref{fig:plot-MfPf}).
Therefore, most of the observational points shown in Fig.~\ref{fig:plot-MfPf} can be reached by modifying the metallicity of the models.
Still, some observational points are located above the theoretical adjustments, not finding agreement even considering the uncertainty in the measured masses.

The He-WD in the PSR~J2043+1711 system ($P = 1.48~\text{d}$) has a mass estimate of 0.173~$M_{\odot}$.
Our model for $M_\text{d,i} = 1.0~M_{\odot}$, $\beta = 0.8$ (green squares in Fig.~\ref{fig:plot-MfPf}) of $P_\text{i} = 7~\text{d}$ has $P_\text{f} = 1.40~\text{d}$ and $M_\text{d,f} = 0.1689~M_{\odot}$, far from the theoretical fits from \citet{Lin2011} and \citet{Tauris1999}, but in great agreement with the observation.

For PSR~J1713+0747 ($P = 67.8~\text{d}$), the millisecond pulsar has 1.35~$M_{\odot}$ and the He-WD companion has 0.292~$M_{\odot}$.
PSR~J1713+0747 is one of the most precisely timed pulsars and provides one of the best pulsar limit on the variation of the gravitational constant, on violations of the universality of free fall, and post-Newtonian parameters measurements \citep{Desvignes2016,Zhu2019}.
Our model for $M_\text{d,i} = 1.2~M_{\odot}$, $\beta = 0.8$ (blue circles in Fig.~\ref{fig:plot-MfPf}) of $P_\text{i} = 30~\text{d}$ has $P_\text{f} = 63.1~\text{d}$ and $M_\text{d,f} = 0.2932~M_{\odot}$, but $M_\text{a,f} = 1.98~M_{\odot}$, not matching the mass of the pulsar.
However, for a rotation plus diffusion sequence with
$M_\text{d,i} = 1.0~M_{\odot}$, $M_\text{a,i} = 1.3~M_{\odot}$, $P_\text{i} = 32~\text{d}$ and $\beta_\text{mt} = 0.9$ we find $M_\text{a,f} = 1.36~M_{\odot}$, $P_\text{f} = 64.7~\text{d}$, and $M_\text{d,f} = 0.2918~M_{\odot}$, in much better agreement with the three measured parameters.

PSR~J1946+3417 ($P = 27.0~\text{d}$) hosts the fourth most massive millisecond pulsar (1.828~$M_{\odot}$), and its He-WD companion has 0.2656~$M_{\odot}$ \citep{Barr2017}.
In this case, taking a rotation plus diffusion sequence with $M_\text{d,i} = 1.0~M_{\odot}$, $M_\text{a,i} = 1.7~M_{\odot}$, $P_\text{i} = 21~\text{d}$, and $\beta_\text{mt} = 0.8$, we find 
$M_\text{a,f} = 1.836~M_{\odot}$, $P_\text{f} = 26.8~\text{d}$, and $M_\text{d,f} = 0.2664~M_{\odot}$, in excellent agreement with the measured parameters.

In summary, the CARB magnetic braking seems to be compatible with the formation of ELM WDs in systems with millisecond pulsars.
The use of low accretion efficiency to the neutrons stars, between 5 and 20 per cent,
revealed to be appropriate, as suggested by \citet{Antoniadis2012,Antoniadis2013,Antoniadis2016}.
Even in cases where the pulsars have masses far from the canonical value ($1.4~M_{\odot}$), a simple adjustment of the initial masses has considerably improved possible matchings of the masses.
Still, we emphasize that different metallicities were not considered up to now.


\subsubsection{Impact of the CARB magnetic braking on the formation of ELM WDs in double degenerates}
Although most of the first ELMs discovered had neutron stars as companions \citep[e.g.,][]{Kerkwijk1996,Kerkwijk2005},
none of the ELMs in the clean sample of the ELM Survey was proven to have a neutron star as a companion \citep{Brown2020ELMVIII}.
Thus, we present preliminary results for models of ELM WDs formed in companion to other white dwarfs.
For WD accretors with $M_\text{a,i}=0.8~M_{\odot}$, we find that there is a systematic shift towards lower final masses, for each given initial period $P_\text{i}$, when compared to the $1.4~M_\odot$ neutron star accretor case.
This result can be easily understood since the evolution of the orbital separation depends on the mass ratio.
Even so, the general behaviour and trends of the
$M_\text{d,f}$--$P_\text{i}$,
$P_\text{i}$--$P_\text{f}$, and
$M_\text{d,f}$--$P_\text{f}$
relations are similar to what we found for neutron star accretors.
Nonetheless, observational data of ELM WDs in double degenerate systems \citep[e.g.,][]{Pelisoli2019GaiaELM,Brown2020ELMVIII}
are widely dispersed in the $M_\text{d,f}$--$P_\text{f}$ plane, moving away from the theoretical low-mass/ELM WDs RLOF models.
This seems to indicate that objects that are not in the region of the quoted theoretical models should have been formed via evolutionary channels other than stable mass transfer, such as common envelope, mergers, triple systems, etc.
Thus, this topic requires further study to say to what extent the disagreement has to do with the quality of the data or with the differences between the parameters of the observed system and the input of the simulations.

\section{Conclusions}
\label{sec:Conclusions}

We first investigated the formation of ELM WDs in binary systems with millisecond pulsars with the proposed
Convection And Rotation Boosted (CARB)
prescription for magnetic braking of \citet{Van2019} and compared their effects on the evolution of LMXB systems with the classic prescription of \citet{Rappaport1983MB}.

We also computed a grid of models of ELM WDs from de ZAMS until a model age of 14~Gyr.
We have considered canonical neutron stars as the accretor, compatible with milliseconds pulsars in LMXBs systems.
Different accretion efficiencies ($\beta_\text{mt}=0.3 ~\text{and}~ 0.8$), and donor initial masses (1.0 and 1.2~$M_{\odot}$) were considered.
The parameter space of the initial orbital period ($P_\text{i}$) was explored between 1 and 300 days, which corresponds to low mass or extremely low mass white dwarf models of final masses between 0.15 and $0.40~M_{\odot}$.
For the evolution of the binary system, we take into account energy loss as gravitational waves, mass loss, spin-orbit coupling, and magnetic braking.
Magnetic braking follows the
CARB prescription \citepalias{Van2019}, which was obtained through a self-consistent deduction considering wind mass loss, rotation, and that the magnetic field is generated due to motions in the convective zone.
Although the CARB model is still considerably simplified --- only radial magnetic fields are considered and the dipole approximation is used, the Alfvenic surface estimated does not depend on the polar angle, the wind considered is isotropic and the rotation axis is assumed aligned to the magnetic field axis --- it has a consistent physical deduction and presents more plausible results when modelling ELM WDs in binary systems.

The use of the CARB magnetic braking prescription by \citetalias{Van2019} strongly modifies the loss of the total angular momentum of the binary systems and, as a natural consequence, also the relation between the orbital period and the white dwarf becomes completely different.
In particular, fine-tuning the initial orbital period is not required.
A range of final masses for the ELM WDs ($0.15\text{--}0.27~M_{\odot}$) can be obtained from a large range of initial orbital periods ($1\text{--}25~\text{d}$), and up to $0.40~M_\odot$ for initial orbital periods up to 300~d.
The bifurcation period ($P_\text{i}=P_\text{f}$) is shifted to longer ones (from 2.75--2.8 to 20--25~days) when the CARB magnetic braking is considered.
I.e., the CARB magnetic braking allows us to get ELM WDs models as light as $0.26~M_{\odot}$ in converging binary systems even with initial orbital periods as long as 20 days,
which is not possible with the empirical magnetic braking prescription of \citetalias{Rappaport1983MB}.
Else, in addition to the LMXB systems, the use of CARB magnetic braking makes it possible to form UCXB systems and also wide-orbit binary millisecond pulsars, which is not possible with the \citetalias{Van2018} prescription.

The orbital period is one of the main factors that relate an ELM WDs as we observe it today with its progenitor system since the initial orbital period is directly linked with the final orbital period and the final mass.
Comparing our models with observational data from He-WDs in binary systems with millisecond pulsars, the use of CARB magnetic braking is shown to be compatible with the formation of ELM WDs in LMXBs.

The main properties of our model grid can be found in Appendix~\ref{sec:Appendix-CompMod}.
In Appendix~\ref{sec:Appendix-PolFits} we provide polynomial fits to the final ELM WD mass ($M_\text{d,f}$) as a function of the initial orbital period ($P_\text{i}$).

\section*{Acknowledgements}

We thank the anonymous referee for the careful reading and helpful suggestions on our work.
We thank Dr. Alejandra Daniela Romero and Dr. Alina Istrate for strong help throughout the research, and
Dr. Ingrid Pelisoli for providing the Gaia DR2 data with reddening corrections.
We thank the Ada IF-UFRGS cluster staff for their support.
This research was developed with the support of the National Supercomputing Center (CESUP), Federal University of Rio Grande do Sul (UFRGS).
This work was carried out with the financial support of the Conselho Nacional de Desenvolvimento Cient\'{i}fico e Tecnol\'{o}gico (CNPq), by the Coordena\c{c}\~{a}o de Aperfei\c{c}oamento de Pessoal de N\'{i}vel Superior - Brasil (CAPES) - Finance Code 001 and FAPERGS.
This research has made extensive use of NASA's Astrophysics Data System.

\section*{Data availability}

The data underlying this article will be shared on reasonable request to the corresponding author.




\bibliographystyle{mnras}
\bibliography{referencias} 

\begin{thebibliography}{}
\makeatletter
\relax
\def\mn@urlcharsother{\let\do\@makeother \do\$\do\&\do\#\do\^\do\_\do\%\do\~}
\def\mn@doi{\begingroup\mn@urlcharsother \@ifnextchar [ {\mn@doi@}
  {\mn@doi@[]}}
\def\mn@doi@[#1]#2{\def\@tempa{#1}\ifx\@tempa\@empty \href
  {http://dx.doi.org/#2} {doi:#2}\else \href {http://dx.doi.org/#2} {#1}\fi
  \endgroup}
\def\mn@eprint#1#2{\mn@eprint@#1:#2::\@nil}
\def\mn@eprint@arXiv#1{\href {http://arxiv.org/abs/#1} {{\tt arXiv:#1}}}
\def\mn@eprint@dblp#1{\href {http://dblp.uni-trier.de/rec/bibtex/#1.xml}
  {dblp:#1}}
\def\mn@eprint@#1:#2:#3:#4\@nil{\def\@tempa {#1}\def\@tempb {#2}\def\@tempc
  {#3}\ifx \@tempc \@empty \let \@tempc \@tempb \let \@tempb \@tempa \fi \ifx
  \@tempb \@empty \def\@tempb {arXiv}\fi \@ifundefined
  {mn@eprint@\@tempb}{\@tempb:\@tempc}{\expandafter \expandafter \csname
  mn@eprint@\@tempb\endcsname \expandafter{\@tempc}}}

\bibitem[\protect\citeauthoryear{{Althaus}, {Panei}, {Romero}, {Rohrmann},
  {C{\'o}rsico}, {Garc{\'\i}a-Berro}  \& {Miller Bertolami}}{{Althaus}
  et~al.}{2009}]{Althaus2009}
{Althaus} L.~G.,  {Panei} J.~A.,  {Romero} A.~D.,  {Rohrmann} R.~D.,
  {C{\'o}rsico} A.~H.,  {Garc{\'\i}a-Berro} E.,   {Miller Bertolami} M.~M.,
  2009, \mn@doi [\aap] {10.1051/0004-6361/200911640}, \href
  {https://ui.adsabs.harvard.edu/abs/2009A&A...502..207A} {502, 207}

\bibitem[\protect\citeauthoryear{{Althaus}, {Miller Bertolami}  \&
  {C{\'o}rsico}}{{Althaus} et~al.}{2013}]{Althaus2013}
{Althaus} L.~G.,  {Miller Bertolami} M.~M.,   {C{\'o}rsico} A.~H.,  2013,
  \mn@doi [\aap] {10.1051/0004-6361/201321868}, \href
  {https://ui.adsabs.harvard.edu/abs/2013A&A...557A..19A} {557, A19}

\bibitem[\protect\citeauthoryear{{Antoniadis}, {van Kerkwijk}, {Koester},
  {Freire}, {Wex}, {Tauris}, {Kramer}  \& {Bassa}}{{Antoniadis}
  et~al.}{2012}]{Antoniadis2012}
{Antoniadis} J.,  {van Kerkwijk} M.~H.,  {Koester} D.,  {Freire} P.~C.~C.,
  {Wex} N.,  {Tauris} T.~M.,  {Kramer} M.,   {Bassa} C.~G.,  2012, \mn@doi
  [\mnras] {10.1111/j.1365-2966.2012.21124.x}, \href
  {https://ui.adsabs.harvard.edu/abs/2012MNRAS.423.3316A} {423, 3316}

\bibitem[\protect\citeauthoryear{{Antoniadis} et~al.,}{{Antoniadis}
  et~al.}{2013}]{Antoniadis2013}
{Antoniadis} J.,  et~al., 2013, \mn@doi [Science] {10.1126/science.1233232},
  \href {https://ui.adsabs.harvard.edu/abs/2013Sci...340..448A} {340, 448}

\bibitem[\protect\citeauthoryear{{Antoniadis}, {Tauris}, {Ozel}, {Barr},
  {Champion}  \& {Freire}}{{Antoniadis} et~al.}{2016}]{Antoniadis2016}
{Antoniadis} J.,  {Tauris} T.~M.,  {Ozel} F.,  {Barr} E.,  {Champion} D.~J.,
  {Freire} P. C.~C.,  2016, arXiv e-prints, \href
  {https://ui.adsabs.harvard.edu/abs/2016arXiv160501665A} {p. arXiv:1605.01665}

\bibitem[\protect\citeauthoryear{{Barr}, {Freire}, {Kramer}, {Champion},
  {Berezina}, {Bassa}, {Lyne}  \& {Stappers}}{{Barr} et~al.}{2017}]{Barr2017}
{Barr} E.~D.,  {Freire} P.~C.~C.,  {Kramer} M.,  {Champion} D.~J.,  {Berezina}
  M.,  {Bassa} C.~G.,  {Lyne} A.~G.,   {Stappers} B.~W.,  2017, \mn@doi
  [\mnras] {10.1093/mnras/stw2947}, \href
  {https://ui.adsabs.harvard.edu/abs/2017MNRAS.465.1711B} {465, 1711}

\bibitem[\protect\citeauthoryear{{Bhattacharya} \& {van den
  Heuvel}}{{Bhattacharya} \& {van den Heuvel}}{1991}]{BhattacharyaHeuvel1991}
{Bhattacharya} D.,  {van den Heuvel} E.~P.~J.,  1991, \mn@doi [\physrep]
  {10.1016/0370-1573(91)90064-S}, \href
  {https://ui.adsabs.harvard.edu/abs/1991PhR...203....1B} {203, 1}

\bibitem[\protect\citeauthoryear{{Bildsten}, {Shen}, {Weinberg}  \&
  {Nelemans}}{{Bildsten} et~al.}{2007}]{BildstenShen2007}
{Bildsten} L.,  {Shen} K.~J.,  {Weinberg} N.~N.,   {Nelemans} G.,  2007,
  \mn@doi [\apjl] {10.1086/519489}, 662, L95

\bibitem[\protect\citeauthoryear{{B{\"o}hm-Vitense}}{{B{\"o}hm-Vitense}}{1958}]{Bohm1958MLT}
{B{\"o}hm-Vitense} E.,  1958, Zeitschrift fur Astrophysik, \href
  {https://ui.adsabs.harvard.edu/abs/1958ZA.....46..108B} {46, 108}

\bibitem[\protect\citeauthoryear{{Breedt}, {G{\"a}nsicke}, {Marsh}, {Steeghs},
  {Drake}  \& {Copperwheat}}{{Breedt} et~al.}{2012}]{BreedtGansickeMarsh2012}
{Breedt} E.,  {G{\"a}nsicke} B.~T.,  {Marsh} T.~R.,  {Steeghs} D.,  {Drake}
  A.~J.,   {Copperwheat} C.~M.,  2012, \mn@doi [\mnras]
  {10.1111/j.1365-2966.2012.21724.x}, 425, 2548

\bibitem[\protect\citeauthoryear{{Brown}, {Kilic}, {Allende Prieto}  \&
  {Kenyon}}{{Brown} et~al.}{2010}]{ELMSurveyI}
{Brown} W.~R.,  {Kilic} M.,  {Allende Prieto} C.,   {Kenyon} S.~J.,  2010,
  \mn@doi [\apj] {10.1088/0004-637X/723/2/1072}, \href
  {https://ui.adsabs.harvard.edu/abs/2010ApJ...723.1072B} {723, 1072}

\bibitem[\protect\citeauthoryear{{Brown}, {Kilic}, {Allende Prieto}  \&
  {Kenyon}}{{Brown} et~al.}{2011}]{BrownKilicPrietoKenyon2011}
{Brown} W.~R.,  {Kilic} M.,  {Allende Prieto} C.,   {Kenyon} S.~J.,  2011,
  \mn@doi [\mnras] {10.1111/j.1745-3933.2010.00986.x}, 411, L31

\bibitem[\protect\citeauthoryear{{Brown}, {Kilic}, {Allende Prieto}  \&
  {Kenyon}}{{Brown} et~al.}{2012}]{ELMSurveyIII}
{Brown} W.~R.,  {Kilic} M.,  {Allende Prieto} C.,   {Kenyon} S.~J.,  2012,
  \mn@doi [\apj] {10.1088/0004-637X/744/2/142}, \href
  {https://ui.adsabs.harvard.edu/abs/2012ApJ...744..142B} {744, 142}

\bibitem[\protect\citeauthoryear{{Brown}, {Kilic}, {Allende Prieto},
  {Gianninas}  \& {Kenyon}}{{Brown} et~al.}{2013}]{ELMSurveyV}
{Brown} W.~R.,  {Kilic} M.,  {Allende Prieto} C.,  {Gianninas} A.,   {Kenyon}
  S.~J.,  2013, \mn@doi [\apj] {10.1088/0004-637X/769/1/66}, \href
  {https://ui.adsabs.harvard.edu/abs/2013ApJ...769...66B} {769, 66}

\bibitem[\protect\citeauthoryear{{Brown}, {Gianninas}, {Kilic}, {Kenyon}  \&
  {Allende Prieto}}{{Brown} et~al.}{2016a}]{ELMSurveyVII}
{Brown} W.~R.,  {Gianninas} A.,  {Kilic} M.,  {Kenyon} S.~J.,   {Allende
  Prieto} C.,  2016a, \mn@doi [\apj] {10.3847/0004-637X/818/2/155}, \href
  {https://ui.adsabs.harvard.edu/abs/2016ApJ...818..155B} {818, 155}

\bibitem[\protect\citeauthoryear{{Brown}, {Kilic}, {Kenyon}  \&
  {Gianninas}}{{Brown} et~al.}{2016b}]{BrownKilicKenyonGianninas2016}
{Brown} W.~R.,  {Kilic} M.,  {Kenyon} S.~J.,   {Gianninas} A.,  2016b, \mn@doi
  [\apj] {10.3847/0004-637x/824/1/46}, 824, 46

\bibitem[\protect\citeauthoryear{{Brown}, {Kilic}  \& {Gianninas}}{{Brown}
  et~al.}{2017}]{BrownKilicGianninas2017}
{Brown} W.~R.,  {Kilic} M.,   {Gianninas} A.,  2017, \mn@doi [\apj]
  {10.3847/1538-4357/aa67e4}, \href
  {https://ui.adsabs.harvard.edu/abs/2017ApJ...839...23B} {839, 23}

\bibitem[\protect\citeauthoryear{{Brown} et~al.,}{{Brown}
  et~al.}{2020a}]{Brown2020ELMVIII}
{Brown} W.~R.,  et~al., 2020a, \mn@doi [\apj] {10.3847/1538-4357/ab63cd}, 889,
  49

\bibitem[\protect\citeauthoryear{Brown, Kilic, B{\'{e}}dard, Kosakowski  \&
  Bergeron}{Brown et~al.}{2020b}]{Brown2020-1201s}
Brown W.~R.,  Kilic M.,  B{\'{e}}dard A.,  Kosakowski A.,   Bergeron P.,
  2020b, \mn@doi [\apj] {10.3847/2041-8213/ab8228}, 892, L35

\bibitem[\protect\citeauthoryear{{Buchler} \& {Yueh}}{{Buchler} \&
  {Yueh}}{1976}]{Buchler1976}
{Buchler} J.~R.,  {Yueh} W.~R.,  1976, \mn@doi [\apj] {10.1086/154847}, \href
  {http://adsabs.harvard.edu/abs/1976ApJ...210..440B} {210, 440}

\bibitem[\protect\citeauthoryear{{Cadelano}, {Ferraro}, {Istrate}, {Pallanca},
  {Lanzoni}  \& {Freire}}{{Cadelano} et~al.}{2019}]{Cadelano2019}
{Cadelano} M.,  {Ferraro} F.~R.,  {Istrate} A.~G.,  {Pallanca} C.,  {Lanzoni}
  B.,   {Freire} P.~C.~C.,  2019, \mn@doi [\apj] {10.3847/1538-4357/ab0e6b},
  \href {https://ui.adsabs.harvard.edu/abs/2019ApJ...875...25C} {875, 25}

\bibitem[\protect\citeauthoryear{{Cassisi}, {Potekhin}, {Pietrinferni},
  {Catelan}  \& {Salaris}}{{Cassisi} et~al.}{2007}]{Cassisi2007}
{Cassisi} S.,  {Potekhin} A.~Y.,  {Pietrinferni} A.,  {Catelan} M.,   {Salaris}
  M.,  2007, \mn@doi [\apj] {10.1086/516819}, \href
  {http://adsabs.harvard.edu/abs/2007ApJ...661.1094C} {661, 1094}

\bibitem[\protect\citeauthoryear{{Chen}, {Han}, {Deca}  \&
  {Podsiadlowski}}{{Chen} et~al.}{2013}]{ChenHanDeca2013}
{Chen} X.,  {Han} Z.,  {Deca} J.,   {Podsiadlowski} P.,  2013, \mn@doi [\mnras]
  {10.1093/mnras/stt992}, \href
  {https://ui.adsabs.harvard.edu/abs/2013MNRAS.434..186C} {434, 186}

\bibitem[\protect\citeauthoryear{{Chen}, {Maxted}, {Li}  \& {Han}}{{Chen}
  et~al.}{2017}]{Chen2017}
{Chen} X.,  {Maxted} P.~F.~L.,  {Li} J.,   {Han} Z.,  2017, \mn@doi [\mnras]
  {10.1093/mnras/stx115}, \href
  {https://ui.adsabs.harvard.edu/abs/2017MNRAS.467.1874C} {467, 1874}

\bibitem[\protect\citeauthoryear{{Chen}, {Tauris}, {Han}  \& {Chen}}{{Chen}
  et~al.}{2021}]{ChenTauris2021}
{Chen} H.-L.,  {Tauris} T.~M.,  {Han} Z.,   {Chen} X.,  2021, \mn@doi [\mnras]
  {10.1093/mnras/stab670}, \href
  {https://ui.adsabs.harvard.edu/abs/2021MNRAS.503.3540C} {503, 3540}

\bibitem[\protect\citeauthoryear{{Chugunov}, {Dewitt}  \&
  {Yakovlev}}{{Chugunov} et~al.}{2007}]{Chugunov2007}
{Chugunov} A.~I.,  {Dewitt} H.~E.,   {Yakovlev} D.~G.,  2007, \mn@doi [\prd]
  {10.1103/PhysRevD.76.025028}, \href
  {https://ui.adsabs.harvard.edu/abs/2007PhRvD..76b5028C} {76, 025028}

\bibitem[\protect\citeauthoryear{{Cyburt} et~al.,}{{Cyburt}
  et~al.}{2010}]{Cyburt2010}
{Cyburt} R.~H.,  et~al., 2010, \mn@doi [\apjs] {10.1088/0067-0049/189/1/240},
  \href {http://adsabs.harvard.edu/abs/2010ApJS..189..240C} {189, 240}

\bibitem[\protect\citeauthoryear{{Deng}, {Li}, {Gao}  \& {Shao}}{{Deng}
  et~al.}{2021}]{Deng2021}
{Deng} Z.-L.,  {Li} X.-D.,  {Gao} Z.-F.,   {Shao} Y.,  2021, \mn@doi [\apj]
  {10.3847/1538-4357/abe0b2}, \href
  {https://ui.adsabs.harvard.edu/abs/2021ApJ...909..174D} {909, 174}

\bibitem[\protect\citeauthoryear{{Desvignes} et~al.,}{{Desvignes}
  et~al.}{2016}]{Desvignes2016}
{Desvignes} G.,  et~al., 2016, \mn@doi [\mnras] {10.1093/mnras/stw483}, \href
  {https://ui.adsabs.harvard.edu/abs/2016MNRAS.458.3341D} {458, 3341}

\bibitem[\protect\citeauthoryear{{Duch{\^e}ne} \& {Kraus}}{{Duch{\^e}ne} \&
  {Kraus}}{2013}]{Duchene2013-ads}
{Duch{\^e}ne} G.,  {Kraus} A.,  2013, \mn@doi [\araa]
  {10.1146/annurev-astro-081710-102602}, \href
  {https://ui.adsabs.harvard.edu/abs/2013ARA&A..51..269D} {51, 269}

\bibitem[\protect\citeauthoryear{{Eggleton}}{{Eggleton}}{1983}]{Eggleton1983}
{Eggleton} P.~P.,  1983, \mn@doi [\apj] {10.1086/160960}, \href
  {http://adsabs.harvard.edu/abs/1983ApJ...268..368E} {268, 368}

\bibitem[\protect\citeauthoryear{{Ferguson}, {Alexander}, {Allard}, {Barman},
  {Bodnarik}, {Hauschildt}, {Heffner-Wong}  \& {Tamanai}}{{Ferguson}
  et~al.}{2005}]{Ferguson2005}
{Ferguson} J.~W.,  {Alexander} D.~R.,  {Allard} F.,  {Barman} T.,  {Bodnarik}
  J.~G.,  {Hauschildt} P.~H.,  {Heffner-Wong} A.,   {Tamanai} A.,  2005,
  \mn@doi [\apj] {10.1086/428642}, \href
  {http://adsabs.harvard.edu/abs/2005ApJ...623..585F} {623, 585}

\bibitem[\protect\citeauthoryear{{Fonseca} et~al.,}{{Fonseca}
  et~al.}{2021}]{Fonseca2021}
{Fonseca} E.,  et~al., 2021, arXiv e-prints, \href
  {https://ui.adsabs.harvard.edu/abs/2021arXiv210400880F} {p. arXiv:2104.00880}

\bibitem[\protect\citeauthoryear{{Fuller}, {Fowler}  \& {Newman}}{{Fuller}
  et~al.}{1985}]{Fuller1985}
{Fuller} G.~M.,  {Fowler} W.~A.,   {Newman} M.~J.,  1985, \mn@doi [\apj]
  {10.1086/163208}, \href {http://adsabs.harvard.edu/abs/1985ApJ...293....1F}
  {293, 1}

\bibitem[\protect\citeauthoryear{{Gianninas}, {Kilic}, {Brown}, {Canton}  \&
  {Kenyon}}{{Gianninas} et~al.}{2015}]{ELMSurveyVI}
{Gianninas} A.,  {Kilic} M.,  {Brown} W.~R.,  {Canton} P.,   {Kenyon} S.~J.,
  2015, \mn@doi [\apj] {10.1088/0004-637X/812/2/167}, \href
  {https://ui.adsabs.harvard.edu/abs/2015ApJ...812..167G} {812, 167}

\bibitem[\protect\citeauthoryear{Han, Podsiadlowski, Maxted  \& Marsh}{Han
  et~al.}{2003}]{Han2003sdB}
Han Z.,  Podsiadlowski P.,  Maxted P. F.~L.,   Marsh T.~R.,  2003, \mn@doi
  [\mnras] {10.1046/j.1365-8711.2003.06451.x}, 341, 669

\bibitem[\protect\citeauthoryear{{Heger}, {Langer}  \& {Woosley}}{{Heger}
  et~al.}{2000}]{Heger2000}
{Heger} A.,  {Langer} N.,   {Woosley} S.~E.,  2000, \mn@doi [\apj]
  {10.1086/308158}, \href
  {https://ui.adsabs.harvard.edu/abs/2000ApJ...528..368H} {528, 368}

\bibitem[\protect\citeauthoryear{{Heger}, {Woosley}  \& {Spruit}}{{Heger}
  et~al.}{2005}]{Heger2005}
{Heger} A.,  {Woosley} S.~E.,   {Spruit} H.~C.,  2005, \mn@doi [\apj]
  {10.1086/429868}, \href
  {https://ui.adsabs.harvard.edu/abs/2005ApJ...626..350H} {626, 350}

\bibitem[\protect\citeauthoryear{{Henyey}, {Vardya}  \& {Bodenheimer}}{{Henyey}
  et~al.}{1965}]{Henyey1965MLT}
{Henyey} L.,  {Vardya} M.~S.,   {Bodenheimer} P.,  1965, \mn@doi [\apj]
  {10.1086/148357}, \href
  {https://ui.adsabs.harvard.edu/abs/1965ApJ...142..841H} {142, 841}

\bibitem[\protect\citeauthoryear{{Hermes} et~al.,}{{Hermes}
  et~al.}{2013a}]{Hermes2013dec}
{Hermes} J.~J.,  et~al., 2013a, \mn@doi [\mnras] {10.1093/mnras/stt1835}, \href
  {https://ui.adsabs.harvard.edu/abs/2013MNRAS.436.3573H} {436, 3573}

\bibitem[\protect\citeauthoryear{{Hermes} et~al.,}{{Hermes}
  et~al.}{2013b}]{Hermes2013mar}
{Hermes} J.~J.,  et~al., 2013b, \mn@doi [\apj] {10.1088/0004-637X/765/2/102},
  \href {https://ui.adsabs.harvard.edu/abs/2013ApJ...765..102H} {765, 102}

\bibitem[\protect\citeauthoryear{{Hurley}, {Tout}  \& {Pols}}{{Hurley}
  et~al.}{2002}]{Hurley2002}
{Hurley} J.~R.,  {Tout} C.~A.,   {Pols} O.~R.,  2002, \mn@doi [\mnras]
  {10.1046/j.1365-8711.2002.05038.x}, \href
  {https://ui.adsabs.harvard.edu/abs/2002MNRAS.329..897H} {329, 897}

\bibitem[\protect\citeauthoryear{{Iben} \& {MacDonald}}{{Iben} \&
  {MacDonald}}{1985}]{Iben1985}
{Iben} I. J.,  {MacDonald} J.,  1985, \mn@doi [\apj] {10.1086/163473}, \href
  {https://ui.adsabs.harvard.edu/abs/1985ApJ...296..540I} {296, 540}

\bibitem[\protect\citeauthoryear{{Iben} \& {Tutukov}}{{Iben} \&
  {Tutukov}}{1984}]{IbenTutukov1984}
{Iben} I.~J.,  {Tutukov} A.~V.,  1984, \mn@doi [\apjs] {10.1086/190932}, \href
  {https://ui.adsabs.harvard.edu/abs/1984ApJS...54..335I} {54, 335}

\bibitem[\protect\citeauthoryear{{Iglesias} \& {Rogers}}{{Iglesias} \&
  {Rogers}}{1993}]{Iglesias1993}
{Iglesias} C.~A.,  {Rogers} F.~J.,  1993, \mn@doi [\apj] {10.1086/172958},
  \href {http://adsabs.harvard.edu/abs/1993ApJ...412..752I} {412, 752}

\bibitem[\protect\citeauthoryear{{Iglesias} \& {Rogers}}{{Iglesias} \&
  {Rogers}}{1996}]{Iglesias1996}
{Iglesias} C.~A.,  {Rogers} F.~J.,  1996, \mn@doi [\apj] {10.1086/177381},
  \href {http://adsabs.harvard.edu/abs/1996ApJ...464..943I} {464, 943}

\bibitem[\protect\citeauthoryear{{Istrate}, {Tauris}  \& {Langer}}{{Istrate}
  et~al.}{2014}]{Istrate2014}
{Istrate} A.~G.,  {Tauris} T.~M.,   {Langer} N.,  2014, \mn@doi [\aap]
  {10.1051/0004-6361/201424680}, \href
  {https://ui.adsabs.harvard.edu/abs/2014A&A...571A..45I} {571, A45}

\bibitem[\protect\citeauthoryear{{Istrate}, {Marchant}, {Tauris}, {Langer},
  {Stancliffe}  \& {Grassitelli}}{{Istrate}
  et~al.}{2016}]{Istrate2016Models-ads}
{Istrate} A.~G.,  {Marchant} P.,  {Tauris} T.~M.,  {Langer} N.,  {Stancliffe}
  R.~J.,   {Grassitelli} L.,  2016, \mn@doi [\aap]
  {10.1051/0004-6361/201628874}, \href
  {https://ui.adsabs.harvard.edu/abs/2016A&A...595A..35I} {595, A35}

\bibitem[\protect\citeauthoryear{{Itoh}, {Hayashi}, {Nishikawa}  \&
  {Kohyama}}{{Itoh} et~al.}{1996}]{Itoh1996a}
{Itoh} N.,  {Hayashi} H.,  {Nishikawa} A.,   {Kohyama} Y.,  1996, \mn@doi
  [\apjs] {10.1086/192264}, \href
  {http://adsabs.harvard.edu/abs/1996ApJS..102..411I} {102, 411}

\bibitem[\protect\citeauthoryear{{Ivanova}}{{Ivanova}}{2006}]{Ivanova2006}
{Ivanova} N.,  2006, \mn@doi [\apjl] {10.1086/510672}, \href
  {https://ui.adsabs.harvard.edu/abs/2006ApJ...653L.137I} {653, L137}

\bibitem[\protect\citeauthoryear{{Joss}, {Rappaport}  \& {Lewis}}{{Joss}
  et~al.}{1987}]{Joss1987}
{Joss} P.~C.,  {Rappaport} S.,   {Lewis} W.,  1987, \mn@doi [\apj]
  {10.1086/165443}, \href
  {https://ui.adsabs.harvard.edu/abs/1987ApJ...319..180J} {319, 180}

\bibitem[\protect\citeauthoryear{{Kawka} \& {Vennes}}{{Kawka} \&
  {Vennes}}{2009}]{Kawka2009}
{Kawka} A.,  {Vennes} S.,  2009, \mn@doi [\aap] {10.1051/0004-6361/200912954},
  \href {https://ui.adsabs.harvard.edu/abs/2009A&A...506L..25K} {506, L25}

\bibitem[\protect\citeauthoryear{{Kawka}, {Simpson}, {Vennes}, {Bessell}, {Da
  Costa}, {Marino}  \& {Murphy}}{{Kawka} et~al.}{2020}]{Kawka2020}
{Kawka} A.,  {Simpson} J.~D.,  {Vennes} S.,  {Bessell} M.~S.,  {Da Costa}
  G.~S.,  {Marino} A.~F.,   {Murphy} S.~J.,  2020, \mn@doi [\mnras]
  {10.1093/mnrasl/slaa068}, \href
  {https://ui.adsabs.harvard.edu/abs/2020MNRAS.495L.129K} {495, L129}

\bibitem[\protect\citeauthoryear{{Kilic}, {Brown}, {Allende Prieto},
  {Pinsonneault}  \& {Kenyon}}{{Kilic} et~al.}{2007}]{Kilic2007}
{Kilic} M.,  {Brown} W.~R.,  {Allende Prieto} C.,  {Pinsonneault} M.~H.,
  {Kenyon} S.~J.,  2007, \mn@doi [\apj] {10.1086/518735}, \href
  {https://ui.adsabs.harvard.edu/abs/2007ApJ...664.1088K} {664, 1088}

\bibitem[\protect\citeauthoryear{{Kilic}, {Brown}, {Allende Prieto},
  {Ag{\"u}eros}, {Heinke}  \& {Kenyon}}{{Kilic} et~al.}{2011}]{ELMSurveyII}
{Kilic} M.,  {Brown} W.~R.,  {Allende Prieto} C.,  {Ag{\"u}eros} M.~A.,
  {Heinke} C.,   {Kenyon} S.~J.,  2011, \mn@doi [\apj]
  {10.1088/0004-637X/727/1/3}, \href
  {https://ui.adsabs.harvard.edu/abs/2011ApJ...727....3K} {727, 3}

\bibitem[\protect\citeauthoryear{{Kilic}, {Brown}, {Allende Prieto}, {Kenyon},
  {Heinke}, {Ag{\"u}eros}  \& {Kleinman}}{{Kilic} et~al.}{2012}]{ELMSurveyIV}
{Kilic} M.,  {Brown} W.~R.,  {Allende Prieto} C.,  {Kenyon} S.~J.,  {Heinke}
  C.~O.,  {Ag{\"u}eros} M.~A.,   {Kleinman} S.~J.,  2012, \mn@doi [\apj]
  {10.1088/0004-637X/751/2/141}, \href
  {https://ui.adsabs.harvard.edu/abs/2012ApJ...751..141K} {751, 141}

\bibitem[\protect\citeauthoryear{{Knigge}, {Baraffe}  \& {Patterson}}{{Knigge}
  et~al.}{2011}]{Knigge2011}
{Knigge} C.,  {Baraffe} I.,   {Patterson} J.,  2011, \mn@doi [\apjs]
  {10.1088/0067-0049/194/2/28}, \href
  {https://ui.adsabs.harvard.edu/abs/2011ApJS..194...28K} {194, 28}

\bibitem[\protect\citeauthoryear{{Kolb} \& {Ritter}}{{Kolb} \&
  {Ritter}}{1990}]{Kolb1990}
{Kolb} U.,  {Ritter} H.,  1990, \aap, \href
  {https://ui.adsabs.harvard.edu/abs/1990A&A...236..385K} {236, 385}

\bibitem[\protect\citeauthoryear{{Korol}, {Rossi}, {Groot}, {Nelemans},
  {Toonen}  \& {Brown}}{{Korol} et~al.}{2017}]{Korol2017}
{Korol} V.,  {Rossi} E.~M.,  {Groot} P.~J.,  {Nelemans} G.,  {Toonen} S.,
  {Brown} A. G.~A.,  2017, \mn@doi [\mnras] {10.1093/mnras/stx1285}, \href
  {https://ui.adsabs.harvard.edu/abs/2017MNRAS.470.1894K} {470, 1894}

\bibitem[\protect\citeauthoryear{{Korol} et~al.,}{{Korol}
  et~al.}{2020}]{Korol2020}
{Korol} V.,  et~al., 2020, \mn@doi [\aap] {10.1051/0004-6361/202037764}, \href
  {https://ui.adsabs.harvard.edu/abs/2020A&A...638A.153K} {638, A153}

\bibitem[\protect\citeauthoryear{{Kosakowski}, {Kilic}, {Brown}  \&
  {Gianninas}}{{Kosakowski} et~al.}{2020}]{Kosakowski2020ELMSSI}
{Kosakowski} A.,  {Kilic} M.,  {Brown} W.~R.,   {Gianninas} A.,  2020, \mn@doi
  [\apj] {10.3847/1538-4357/ab8300}, \href
  {https://ui.adsabs.harvard.edu/abs/2020ApJ...894...53K} {894, 53}

\bibitem[\protect\citeauthoryear{{Kraft}, {Mathews}  \& {Greenstein}}{{Kraft}
  et~al.}{1962}]{Kraft1962}
{Kraft} R.~P.,  {Mathews} J.,   {Greenstein} J.~L.,  1962, \mn@doi [\apj]
  {10.1086/147381}, \href
  {https://ui.adsabs.harvard.edu/abs/1962ApJ...136..312K} {136, 312}

\bibitem[\protect\citeauthoryear{{Kulkarni} \& {van Kerkwijk}}{{Kulkarni} \&
  {van Kerkwijk}}{2010}]{Kulkarni2010}
{Kulkarni} S.~R.,  {van Kerkwijk} M.~H.,  2010, \mn@doi [\apj]
  {10.1088/0004-637X/719/2/1123}, \href
  {https://ui.adsabs.harvard.edu/abs/2010ApJ...719.1123K} {719, 1123}

\bibitem[\protect\citeauthoryear{{Kupfer} et~al.,}{{Kupfer}
  et~al.}{2015}]{Kupfer2015}
{Kupfer} T.,  et~al., 2015, \mn@doi [\aap] {10.1051/0004-6361/201425213}, \href
  {https://ui.adsabs.harvard.edu/abs/2015A&A...576A..44K} {576, A44}

\bibitem[\protect\citeauthoryear{{Kupfer} et~al.,}{{Kupfer}
  et~al.}{2018}]{Kupfer2018}
{Kupfer} T.,  et~al., 2018, \mn@doi [\mnras] {10.1093/mnras/sty1545}, \href
  {https://ui.adsabs.harvard.edu/abs/2018MNRAS.480..302K} {480, 302}

\bibitem[\protect\citeauthoryear{{Lamberts}, {Blunt}, {Littenberg},
  {Garrison-Kimmel}, {Kupfer}  \& {Sanderson}}{{Lamberts}
  et~al.}{2019}]{Lamberts2019}
{Lamberts} A.,  {Blunt} S.,  {Littenberg} T.~B.,  {Garrison-Kimmel} S.,
  {Kupfer} T.,   {Sanderson} R.~E.,  2019, \mn@doi [\mnras]
  {10.1093/mnras/stz2834}, \href
  {https://ui.adsabs.harvard.edu/abs/2019MNRAS.490.5888L} {490, 5888}

\bibitem[\protect\citeauthoryear{{Landau} \& {Lifshitz}}{{Landau} \&
  {Lifshitz}}{1975}]{Landau1975CTF}
{Landau} L.~D.,  {Lifshitz} E.~M.,  1975, {The classical theory of fields}.
Pergamon Press, Oxford

\bibitem[\protect\citeauthoryear{{Langanke} \&
  {Mart{\'{\i}}nez-Pinedo}}{{Langanke} \&
  {Mart{\'{\i}}nez-Pinedo}}{2000}]{Langanke2000}
{Langanke} K.,  {Mart{\'{\i}}nez-Pinedo} G.,  2000, \mn@doi [\nphysa]
  {10.1016/S0375-9474(00)00131-7}, \href
  {http://adsabs.harvard.edu/abs/2000NuPhA.673..481L} {673, 481}

\bibitem[\protect\citeauthoryear{{Li}, {Chen}, {Chen}  \& {Han}}{{Li}
  et~al.}{2019}]{LiChenHan2019}
{Li} Z.,  {Chen} X.,  {Chen} H.-L.,   {Han} Z.,  2019, \mn@doi [\apj]
  {10.3847/1538-4357/aaf9a1}, \href
  {https://ui.adsabs.harvard.edu/abs/2019ApJ...871..148L} {871, 148}

\bibitem[\protect\citeauthoryear{{Li}, {Chen}, {Chen}, {Li}, {Yu}  \&
  {Han}}{{Li} et~al.}{2020}]{LiChenChenLiYuHan2020}
{Li} Z.,  {Chen} X.,  {Chen} H.-L.,  {Li} J.,  {Yu} S.,   {Han} Z.,  2020,
  \mn@doi [\apj] {10.3847/1538-4357/ab7dc2}, \href
  {https://ui.adsabs.harvard.edu/abs/2020ApJ...893....2L} {893, 2}

\bibitem[\protect\citeauthoryear{{Lin}, {Rappaport}, {Podsiadlowski}, {Nelson},
  {Paxton}  \& {Todorov}}{{Lin} et~al.}{2011}]{Lin2011}
{Lin} J.,  {Rappaport} S.,  {Podsiadlowski} P.,  {Nelson} L.,  {Paxton} B.,
  {Todorov} P.,  2011, \mn@doi [\apj] {10.1088/0004-637X/732/2/70}, \href
  {https://ui.adsabs.harvard.edu/abs/2011ApJ...732...70L} {732, 70}

\bibitem[\protect\citeauthoryear{{Liu} et~al.,}{{Liu}
  et~al.}{2020}]{LiuIstrate2020}
{Liu} K.,  et~al., 2020, \mn@doi [\mnras] {10.1093/mnras/staa2993}, \href
  {https://ui.adsabs.harvard.edu/abs/2020MNRAS.499.2276L} {499, 2276}

\bibitem[\protect\citeauthoryear{{Ma} \& {Li}}{{Ma} \& {Li}}{2009}]{MaLi2009}
{Ma} B.,  {Li} X.-D.,  2009, \mn@doi [\apj] {10.1088/0004-637X/691/2/1611},
  \href {https://ui.adsabs.harvard.edu/abs/2009ApJ...691.1611M} {691, 1611}

\bibitem[\protect\citeauthoryear{{Mata S{\'a}nchez}, {Istrate}, {van Kerkwijk},
  {Breton}  \& {Kaplan}}{{Mata S{\'a}nchez} et~al.}{2020}]{Sanchez2020}
{Mata S{\'a}nchez} D.,  {Istrate} A.~G.,  {van Kerkwijk} M.~H.,  {Breton}
  R.~P.,   {Kaplan} D.~L.,  2020, \mn@doi [\mnras] {10.1093/mnras/staa983},
  \href {https://ui.adsabs.harvard.edu/abs/2020MNRAS.494.4031M} {494, 4031}

\bibitem[\protect\citeauthoryear{{Maxted} et~al.,}{{Maxted}
  et~al.}{2014}]{Maxted2014}
{Maxted} P.~F.~L.,  et~al., 2014, \mn@doi [\mnras] {10.1093/mnras/stt2007},
  \href {https://ui.adsabs.harvard.edu/abs/2014MNRAS.437.1681M} {437, 1681}

\bibitem[\protect\citeauthoryear{{Mestel}}{{Mestel}}{1968}]{Mestel1968}
{Mestel} L.,  1968, \mn@doi [\mnras] {10.1093/mnras/138.3.359}, \href
  {https://ui.adsabs.harvard.edu/abs/1968MNRAS.138..359M} {138, 359}

\bibitem[\protect\citeauthoryear{{Mestel} \& {Spruit}}{{Mestel} \&
  {Spruit}}{1987}]{MestelSpruit1987}
{Mestel} L.,  {Spruit} H.~C.,  1987, \mn@doi [\mnras] {10.1093/mnras/226.1.57},
  \href {https://ui.adsabs.harvard.edu/abs/1987MNRAS.226...57M} {226, 57}

\bibitem[\protect\citeauthoryear{{Moore}, {Cole}  \& {Berry}}{{Moore}
  et~al.}{2015}]{Moore2015}
{Moore} C.~J.,  {Cole} R.~H.,   {Berry} C.~P.~L.,  2015, \mn@doi [Classical and
  Quantum Gravity] {10.1088/0264-9381/32/1/015014}, \href
  {https://ui.adsabs.harvard.edu/abs/2015CQGra..32a5014M} {32, 015014}

\bibitem[\protect\citeauthoryear{{Oda}, {Hino}, {Muto}, {Takahara}  \&
  {Sato}}{{Oda} et~al.}{1994}]{Oda1994}
{Oda} T.,  {Hino} M.,  {Muto} K.,  {Takahara} M.,   {Sato} K.,  1994, \mn@doi
  [Atomic Data and Nuclear Data Tables] {10.1006/adnd.1994.1007}, \href
  {http://adsabs.harvard.edu/abs/1994ADNDT..56..231O} {56, 231}

\bibitem[\protect\citeauthoryear{{{\"O}zel} \& {Freire}}{{{\"O}zel} \&
  {Freire}}{2016}]{Ozel2016}
{{\"O}zel} F.,  {Freire} P.,  2016, \mn@doi [\araa]
  {10.1146/annurev-astro-081915-023322}, \href
  {https://ui.adsabs.harvard.edu/abs/2016ARA&A..54..401O} {54, 401}

\bibitem[\protect\citeauthoryear{{Panei}, {Althaus}, {Chen}  \& {Han}}{{Panei}
  et~al.}{2007}]{Panei2007}
{Panei} J.~A.,  {Althaus} L.~G.,  {Chen} X.,   {Han} Z.,  2007, \mn@doi
  [\mnras] {10.1111/j.1365-2966.2007.12400.x}, \href
  {https://ui.adsabs.harvard.edu/abs/2007MNRAS.382..779P} {382, 779}

\bibitem[\protect\citeauthoryear{{Patterson}}{{Patterson}}{1984}]{Patterson1984}
{Patterson} J.,  1984, \mn@doi [\apjs] {10.1086/190940}, \href
  {https://ui.adsabs.harvard.edu/abs/1984ApJS...54..443P} {54, 443}

\bibitem[\protect\citeauthoryear{{Paxton}, {Bildsten}, {Dotter}, {Herwig},
  {Lesaffre}  \& {Timmes}}{{Paxton} et~al.}{2011}]{Paxton2011}
{Paxton} B.,  {Bildsten} L.,  {Dotter} A.,  {Herwig} F.,  {Lesaffre} P.,
  {Timmes} F.,  2011, \mn@doi [\apjs] {10.1088/0067-0049/192/1/3}, \href
  {http://adsabs.harvard.edu/abs/2011ApJS..192....3P} {192, 3}

\bibitem[\protect\citeauthoryear{{Paxton} et~al.,}{{Paxton}
  et~al.}{2013}]{Paxton2013}
{Paxton} B.,  et~al., 2013, \mn@doi [\apjs] {10.1088/0067-0049/208/1/4}, \href
  {http://adsabs.harvard.edu/abs/2013ApJS..208....4P} {208, 4}

\bibitem[\protect\citeauthoryear{{Paxton} et~al.,}{{Paxton}
  et~al.}{2015}]{Paxton2015}
{Paxton} B.,  et~al., 2015, \mn@doi [\apjs] {10.1088/0067-0049/220/1/15}, \href
  {http://adsabs.harvard.edu/abs/2015ApJS..220...15P} {220, 15}

\bibitem[\protect\citeauthoryear{{Paxton} et~al.,}{{Paxton}
  et~al.}{2018}]{Paxton2018}
{Paxton} B.,  et~al., 2018, \mn@doi [\apjs] {10.3847/1538-4365/aaa5a8}, \href
  {http://adsabs.harvard.edu/abs/2018ApJS..234...34P} {234, 34}

\bibitem[\protect\citeauthoryear{{Paxton} et~al.,}{{Paxton}
  et~al.}{2019}]{Paxton2019}
{Paxton} B.,  et~al., 2019, \mn@doi [\apjs] {10.3847/1538-4365/ab2241}, \href
  {https://ui.adsabs.harvard.edu/abs/2019ApJS..243...10P} {243, 10}

\bibitem[\protect\citeauthoryear{{Pelisoli} \& {Vos}}{{Pelisoli} \&
  {Vos}}{2019}]{Pelisoli2019GaiaELM}
{Pelisoli} I.,  {Vos} J.,  2019, \mn@doi [\mnras] {10.1093/mnras/stz1876},
  \href {https://ui.adsabs.harvard.edu/abs/2019MNRAS.488.2892P} {488, 2892}

\bibitem[\protect\citeauthoryear{{Podsiadlowski}, {Rappaport}  \&
  {Pfahl}}{{Podsiadlowski} et~al.}{2002}]{Podsiadlowski2002}
{Podsiadlowski} P.,  {Rappaport} S.,   {Pfahl} E.~D.,  2002, \mn@doi [\apj]
  {10.1086/324686}, \href
  {https://ui.adsabs.harvard.edu/abs/2002ApJ...565.1107P} {565, 1107}

\bibitem[\protect\citeauthoryear{{Pols}, {Tout}, {Eggleton}  \& {Han}}{{Pols}
  et~al.}{1995}]{Pols1995}
{Pols} O.~R.,  {Tout} C.~A.,  {Eggleton} P.~P.,   {Han} Z.,  1995, \mn@doi
  [\mnras] {10.1093/mnras/274.3.964}, \href
  {http://adsabs.harvard.edu/abs/1995MNRAS.274..964P} {274, 964}

\bibitem[\protect\citeauthoryear{{Potekhin} \& {Chabrier}}{{Potekhin} \&
  {Chabrier}}{2010}]{Potekhin2010}
{Potekhin} A.~Y.,  {Chabrier} G.,  2010, \mn@doi [Contributions to Plasma
  Physics] {10.1002/ctpp.201010017}, \href
  {http://adsabs.harvard.edu/abs/2010CoPP...50...82P} {50, 82}

\bibitem[\protect\citeauthoryear{{Pylyser} \& {Savonije}}{{Pylyser} \&
  {Savonije}}{1988}]{PylyserSavonije1988}
{Pylyser} E.,  {Savonije} G.~J.,  1988, \aap, \href
  {https://ui.adsabs.harvard.edu/abs/1988A&A...191...57P} {191, 57}

\bibitem[\protect\citeauthoryear{{Pylyser} \& {Savonije}}{{Pylyser} \&
  {Savonije}}{1989}]{PylyserSavonije1989}
{Pylyser} E.~H.~P.,  {Savonije} G.~J.,  1989, \aap, \href
  {https://ui.adsabs.harvard.edu/abs/1989A&A...208...52P} {208, 52}

\bibitem[\protect\citeauthoryear{{Rappaport}, {Verbunt}  \& {Joss}}{{Rappaport}
  et~al.}{1983}]{Rappaport1983MB}
{Rappaport} S.,  {Verbunt} F.,   {Joss} P.~C.,  1983, \mn@doi [\apj]
  {10.1086/161569}, \href
  {https://ui.adsabs.harvard.edu/abs/1983ApJ...275..713R} {275, 713}

\bibitem[\protect\citeauthoryear{{Rappaport}, {Podsiadlowski}, {Joss}, {Di
  Stefano}  \& {Han}}{{Rappaport} et~al.}{1995}]{Rappaport1995}
{Rappaport} S.,  {Podsiadlowski} P.,  {Joss} P.~C.,  {Di Stefano} R.,   {Han}
  Z.,  1995, \mn@doi [\mnras] {10.1093/mnras/273.3.731}, \href
  {https://ui.adsabs.harvard.edu/abs/1995MNRAS.273..731R} {273, 731}

\bibitem[\protect\citeauthoryear{{Refsdal} \& {Weigert}}{{Refsdal} \&
  {Weigert}}{1971}]{Refsdal1971}
{Refsdal} S.,  {Weigert} A.,  1971, \aap, \href
  {https://ui.adsabs.harvard.edu/abs/1971A&A....13..367R} {13, 367}

\bibitem[\protect\citeauthoryear{{Reimers}}{{Reimers}}{1975}]{Reimers1975}
{Reimers} D.,  1975, {Circumstellar envelopes and mass loss of red giant
  stars.}.
Springer, New York, pp 229--256

\bibitem[\protect\citeauthoryear{{R{\'e}ville}, {Brun}, {Matt}, {Strugarek}  \&
  {Pinto}}{{R{\'e}ville} et~al.}{2015}]{Reville2015}
{R{\'e}ville} V.,  {Brun} A.~S.,  {Matt} S.~P.,  {Strugarek} A.,   {Pinto}
  R.~F.,  2015, \mn@doi [\apj] {10.1088/0004-637x/798/2/116}, 798, 116

\bibitem[\protect\citeauthoryear{{Rogers} \& {Nayfonov}}{{Rogers} \&
  {Nayfonov}}{2002}]{Rogers2002}
{Rogers} F.~J.,  {Nayfonov} A.,  2002, \mn@doi [\apj] {10.1086/341894}, \href
  {http://adsabs.harvard.edu/abs/2002ApJ...576.1064R} {576, 1064}

\bibitem[\protect\citeauthoryear{{Rohrmann}, {Althaus}  \& {Kepler}}{{Rohrmann}
  et~al.}{2011}]{Rohrmann2011}
{Rohrmann} R.~D.,  {Althaus} L.~G.,   {Kepler} S.~O.,  2011, \mn@doi [\mnras]
  {10.1111/j.1365-2966.2010.17716.x}, \href
  {https://ui.adsabs.harvard.edu/abs/2011MNRAS.411..781R} {411, 781}

\bibitem[\protect\citeauthoryear{{Salaris} \& {Cassisi}}{{Salaris} \&
  {Cassisi}}{2017}]{SalarisCassisi2017}
{Salaris} M.,  {Cassisi} S.,  2017, \mn@doi [Royal Society Open Science]
  {10.1098/rsos.170192}, \href
  {https://ui.adsabs.harvard.edu/abs/2017RSOS....470192S} {4, 170192}

\bibitem[\protect\citeauthoryear{{Saumon}, {Chabrier}  \& {van Horn}}{{Saumon}
  et~al.}{1995}]{Saumon1995}
{Saumon} D.,  {Chabrier} G.,   {van Horn} H.~M.,  1995, \mn@doi [\apjs]
  {10.1086/192204}, \href {http://adsabs.harvard.edu/abs/1995ApJS...99..713S}
  {99, 713}

\bibitem[\protect\citeauthoryear{{Schatzman}}{{Schatzman}}{1962}]{Schatzman1962}
{Schatzman} E.,  1962, Annales d'Astrophysique, \href
  {https://ui.adsabs.harvard.edu/abs/1962AnAp...25...18S} {25, 18}

\bibitem[\protect\citeauthoryear{{Serenelli}, {Althaus}, {Rohrmann}  \&
  {Benvenuto}}{{Serenelli} et~al.}{2002}]{Serenelli2002}
{Serenelli} A.~M.,  {Althaus} L.~G.,  {Rohrmann} R.~D.,   {Benvenuto} O.~G.,
  2002, \mn@doi [\mnras] {10.1046/j.1365-8711.2002.05994.x}, \href
  {https://ui.adsabs.harvard.edu/abs/2002MNRAS.337.1091S} {337, 1091}

\bibitem[\protect\citeauthoryear{{Shao} \& {Li}}{{Shao} \&
  {Li}}{2015}]{Shao2015}
{Shao} Y.,  {Li} X.-D.,  2015, \mn@doi [\apj] {10.1088/0004-637X/809/1/99},
  \href {https://ui.adsabs.harvard.edu/abs/2015ApJ...809...99S} {809, 99}

\bibitem[\protect\citeauthoryear{{Skumanich}}{{Skumanich}}{1972}]{Skumanich1972}
{Skumanich} A.,  1972, \mn@doi [\apj] {10.1086/151310}, \href
  {https://ui.adsabs.harvard.edu/abs/1972ApJ...171..565S} {171, 565}

\bibitem[\protect\citeauthoryear{{Smith}}{{Smith}}{1979}]{Smith1979}
{Smith} M.~A.,  1979, \mn@doi [\pasp] {10.1086/130579}, \href
  {https://ui.adsabs.harvard.edu/abs/1979PASP...91..737S} {91, 737}

\bibitem[\protect\citeauthoryear{{Soberman}, {Phinney}  \& {van den
  Heuvel}}{{Soberman} et~al.}{1997}]{Soberman1997}
{Soberman} G.~E.,  {Phinney} E.~S.,   {van den Heuvel} E.~P.~J.,  1997, \aap,
  \href {https://ui.adsabs.harvard.edu/abs/1997A&A...327..620S} {327, 620}

\bibitem[\protect\citeauthoryear{Sun \& Arras}{Sun \&
  Arras}{2018}]{Sun2018Formation}
Sun M.,  Arras P.,  2018, \mn@doi [\apj] {10.3847/1538-4357/aab9a4}, 858, 14

\bibitem[\protect\citeauthoryear{{Tauris}}{{Tauris}}{2018}]{Tauris2018}
{Tauris} T.~M.,  2018, \mn@doi [\prl] {10.1103/PhysRevLett.121.131105}, \href
  {https://ui.adsabs.harvard.edu/abs/2018PhRvL.121m1105T} {121, 131105}

\bibitem[\protect\citeauthoryear{{Tauris} \& {Savonije}}{{Tauris} \&
  {Savonije}}{1999}]{Tauris1999}
{Tauris} T.~M.,  {Savonije} G.~J.,  1999, \aap, \href
  {https://ui.adsabs.harvard.edu/abs/1999A&A...350..928T} {350, 928}

\bibitem[\protect\citeauthoryear{{Tauris} \& {van den Heuvel}}{{Tauris} \& {van
  den Heuvel}}{2006}]{TaurisHeuvel2006}
{Tauris} T.~M.,  {van den Heuvel} E.~P.~J.,  2006, {Formation and evolution of
  compact stellar X-ray sources}.
Cambridge University Press, pp 623--665

\bibitem[\protect\citeauthoryear{{Tauris}, {Langer}  \& {Kramer}}{{Tauris}
  et~al.}{2012}]{TaurisLangerKramer2012}
{Tauris} T.~M.,  {Langer} N.,   {Kramer} M.,  2012, \mn@doi [\mnras]
  {10.1111/j.1365-2966.2012.21446.x}, \href
  {https://ui.adsabs.harvard.edu/abs/2012MNRAS.425.1601T} {425, 1601}

\bibitem[\protect\citeauthoryear{{Thoul}, {Bahcall}  \& {Loeb}}{{Thoul}
  et~al.}{1994}]{Thoul1994}
{Thoul} A.~A.,  {Bahcall} J.~N.,   {Loeb} A.,  1994, \mn@doi [\apj]
  {10.1086/173695}, \href
  {https://ui.adsabs.harvard.edu/abs/1994ApJ...421..828T} {421, 828}

\bibitem[\protect\citeauthoryear{{Timmes} \& {Swesty}}{{Timmes} \&
  {Swesty}}{2000}]{Timmes2000}
{Timmes} F.~X.,  {Swesty} F.~D.,  2000, \mn@doi [\apjs] {10.1086/313304}, \href
  {http://adsabs.harvard.edu/abs/2000ApJS..126..501T} {126, 501}

\bibitem[\protect\citeauthoryear{{Toloza} et~al.,}{{Toloza}
  et~al.}{2019}]{Toloza2019}
{Toloza} O.,  et~al., 2019, \baas, \href
  {https://ui.adsabs.harvard.edu/abs/2019BAAS...51c.168T} {51, 168}

\bibitem[\protect\citeauthoryear{{Tutukov} \& {Cherepashchuk}}{{Tutukov} \&
  {Cherepashchuk}}{2020}]{Tutukov2020}
{Tutukov} A.~V.,  {Cherepashchuk} A.~M.,  2020, \mn@doi [Physics Uspekhi]
  {10.3367/UFNe.2019.03.038547}, \href
  {https://ui.adsabs.harvard.edu/abs/2020PhyU...63..209T} {63, 209}

\bibitem[\protect\citeauthoryear{{Van} \& {Ivanova}}{{Van} \&
  {Ivanova}}{2019}]{Van2019}
{Van} K.~X.,  {Ivanova} N.,  2019, \mn@doi [\apj] {10.3847/2041-8213/ab571c},
  886, L31

\bibitem[\protect\citeauthoryear{Van, Ivanova  \& Heinke}{Van
  et~al.}{2018}]{Van2018}
Van K.~X.,  Ivanova N.,   Heinke C.~O.,  2018, \mn@doi [\mnras]
  {10.1093/mnras/sty3489}, 483, 5595

\bibitem[\protect\citeauthoryear{{Webbink}}{{Webbink}}{1984}]{Webbink1984}
{Webbink} R.~F.,  1984, \mn@doi [\apj] {10.1086/161701}, \href
  {https://ui.adsabs.harvard.edu/abs/1984ApJ...277..355W} {277, 355}

\bibitem[\protect\citeauthoryear{{Webbink}, {Rappaport}  \&
  {Savonije}}{{Webbink} et~al.}{1983}]{Webbink1983}
{Webbink} R.~F.,  {Rappaport} S.,   {Savonije} G.~J.,  1983, \mn@doi [\apj]
  {10.1086/161159}, \href
  {https://ui.adsabs.harvard.edu/abs/1983ApJ...270..678W} {270, 678}

\bibitem[\protect\citeauthoryear{{Weber} \& {Davis}}{{Weber} \&
  {Davis}}{1967}]{Weber1967}
{Weber} E.~J.,  {Davis} Leverett J.,  1967, \mn@doi [\apj] {10.1086/149138},
  \href {https://ui.adsabs.harvard.edu/abs/1967ApJ...148..217W} {148, 217}

\bibitem[\protect\citeauthoryear{{Zhang}, {Jeffery}, {Chen}  \& {Han}}{{Zhang}
  et~al.}{2014}]{ZhangJefferyChenHan2014}
{Zhang} X.,  {Jeffery} C.~S.,  {Chen} X.,   {Han} Z.,  2014, \mn@doi [\mnras]
  {10.1093/mnras/stu1741}, 445, 660

\bibitem[\protect\citeauthoryear{{Zhu} et~al.,}{{Zhu} et~al.}{2019}]{Zhu2019}
{Zhu} W.~W.,  et~al., 2019, \mn@doi [\mnras] {10.1093/mnras/sty2905}, \href
  {https://ui.adsabs.harvard.edu/abs/2019MNRAS.482.3249Z} {482, 3249}

\bibitem[\protect\citeauthoryear{{van Kerkwijk}, {Bergeron}  \&
  {Kulkarni}}{{van Kerkwijk} et~al.}{1996}]{Kerkwijk1996}
{van Kerkwijk} M.~H.,  {Bergeron} P.,   {Kulkarni} S.~R.,  1996, \mn@doi
  [\apjl] {10.1086/310209}, \href
  {https://ui.adsabs.harvard.edu/abs/1996ApJ...467L..89V} {467, L89}

\bibitem[\protect\citeauthoryear{{van Kerkwijk}, {Bassa}, {Jacoby}  \&
  {Jonker}}{{van Kerkwijk} et~al.}{2005}]{Kerkwijk2005}
{van Kerkwijk} M.~H.,  {Bassa} C.~G.,  {Jacoby} B.~A.,   {Jonker} P.~G.,  2005,
  in {Rasio} F.~A.,  {Stairs} I.~H.,  eds,  Astronomical Society of the Pacific
  Conference Series Vol. 328, Binary Radio Pulsars. p.~357 (\mn@eprint {arXiv}
  {astro-ph/0405283})

\makeatother
\end{thebibliography}




\appendix

\section{Computed models}
\label{sec:Appendix-CompMod}

In Tables~\ref{tab:M1014Z02-rot},
\ref{tab:M1214Z02-rot}, and
\ref{tab:M1014Z02-dif}
we present the properties of the computed models at final age (14~Gyr).
These three tables correspond to the initial setups \# 1, 2 and 3 in Table~\ref{tab:setups}, respectively.
The CARB magnetic braking prescription \citep{Van2019} is considered in all cases.
For each model we show the initial and final orbital periods, the final masses of the donor and the accreting stars, and the number of hydrogen shell flashes that occurred during evolution.
More details are given in the caption of each table.

\begin{table*}
\begin{minipage}{140mm}
	\centering
	\caption{Grid of ELM WDs models considering the use of the enhanced CARB magnetic braking prescription proposed by \citet{Van2019} in the angular momentum evolution of the LMXB systems for various initial orbital periods (third column). The first and second columns show the final donor and accretor masses for each simulation after 14~Gyr, respectively. The fourth column indicates the number of hydrogen shell flashes (\#HSF). The fifth column is the frequency of the gravitational wave emitted by the binaries. The sixth and the seventh columns are the dimensionless gravitational wave amplitude and the characteristic strain, respectively. The last column is the merging time due to gravitational radiation. Details on how we calculate the values on the last four columns are provided in the Appendix~\ref{sec:Appendix-GW}. $M_{\text{d,i}}=1.0~M_{\odot}$, $M_{\text{a,i}}=1.4~M_{\odot}$, $Z=0.02$, and $\beta_\text{mt}=0.3$ for all models. Rotation is taking into account in all models.}
	\label{tab:M1014Z02-rot}
	\begin{tabular}{cccccccc}
\hline
$M_{\text{d,f}}~(M_{\odot})$ &
$M_{\text{a,f}}~(M_{\odot})$ &
$P_\text{i} / P_\text{f}~(\text{d})$ &
\#HSF &
$f_\text{GW}~\text{(Hz)}$ &
$\mathcal{A}$ &
$h_\text{c}$ &
$\tau_\text{GW}~\text{(Myr)}$ \\
\hline
0.4067 & 1.594 & 300/809 & 0  \\
0.3894 & 1.645 & 200/609 & 0  \\
0.3775 & 1.677 & 150/492 & 0  \\
0.3609 & 1.716 & 100/355 & 0  \\
0.3455 & 1.749 & 70/254 & 0  \\
0.3286 & 1.783 & 50/169 & 0  \\
0.3146 & 1.808 & 40/116 & 0  \\
0.2886 & 1.850 & 30/55.3 & 1  \\
0.2730 & 1.873 & 25/32.1 & 1  \\
0.2577 & 1.891 & 20/18.2 & 1  \\
0.2530 & 1.896 & 19/15.2 & 1  \\
0.2482 & 1.902 & 18/12.6 & 1  \\
0.2433 & 1.907 & 17/10.4 & 1  \\
0.2383 & 1.913 & 16/8.62 & 1  \\
0.2290 & 1.923 & 15/7.58 & 2  \\
0.2218 & 1.931 & 14/6.28 & 2  \\
0.2145 & 1.938 & 13/5.10 & 2  \\
0.2090 & 1.943 & 12/4.01 & 0  \\
0.2032 & 1.948 & 11/3.18 & 0  \\
0.1979 & 1.952 & 10/2.50 & 0  \\
0.1933 & 1.956 & 9/1.92 & 0 & $1.21 \times 10^{-5}$ & $9.15 \times 10^{-24}$ & $3.57 \times 10^{-22}$ & $9.15 \times 10^{5}$ \\
0.1883 & 1.960 & 8/1.50 & 0 & $1.54 \times 10^{-5}$ & $1.05 \times 10^{-23}$ & $4.65 \times 10^{-22}$ & $4.86 \times 10^{5}$ \\
0.1834 & 1.964 & 7/1.17 & 0 & $1.98 \times 10^{-5}$ & $1.21 \times 10^{-23}$ & $6.06 \times 10^{-22}$ & $2.56 \times 10^{5}$ \\
0.1783 & 1.968 & 6/0.90 & 0 & $2.57 \times 10^{-5}$ & $1.41 \times 10^{-23}$ & $8.02 \times 10^{-22}$ & $1.31 \times 10^{5}$ \\
0.1725 & 1.973 & 5/0.66 & 0 & $3.51 \times 10^{-5}$ & $1.68 \times 10^{-23}$ & $1.12 \times 10^{-21}$ & $5.89 \times 10^{4}$ \\
0.1636 & 1.980 & 4/0.41 & 0 & $5.65 \times 10^{-5}$ & $2.20 \times 10^{-23}$ & $1.85 \times 10^{-21}$ & $1.74 \times 10^{4}$ \\
0.1547 & 1.986 & 3.5/0.24 & 0 & $9.65 \times 10^{-5}$ & $2.98 \times 10^{-23}$ & $3.29 \times 10^{-21}$ & $4.40 \times 10^{3}$ \\
0.1456 & 1.993 & 3.25/0.082 & 0 & $2.82 \times 10^{-4}$ & $5.76 \times 10^{-23}$ & $1.09 \times 10^{-20}$ & $2.65 \times 10^{2}$ \\
\hline
	\end{tabular}
\end{minipage}
\end{table*}
%

\begin{table*}
\begin{minipage}{140mm}
	\centering
	\caption{Grid of ELM WDs models considering the use of the enhanced CARB magnetic braking prescription proposed by \citet{Van2019}, as in Table~\ref{tab:M1014Z02-rot} but for $M_{\text{d,i}}=1.2~M_{\odot}$.}
	\label{tab:M1214Z02-rot}
	\begin{tabular}{cccccccc}
\hline
$M_{\text{d,f}}~(M_{\odot})$ &
$M_{\text{a,f}}~(M_{\odot})$ &
$P_\text{i} / P_\text{f}~(\text{d})$ &
\#HSF &
$f_\text{GW}~\text{(Hz)}$ &
$\mathcal{A}$ &
$h_\text{c}$ &
$\tau_\text{GW}~\text{(Myr)}$ \\
\hline
0.4212 & 1.716 & 300/994 & 0  \\
0.4016 & 1.769 & 200/737 & 0  \\
0.3885 & 1.802 & 150/591 & 0  \\
0.3705 & 1.842 & 100/424 & 0  \\
0.3538 & 1.877 & 70/301 & 0  \\
0.3357 & 1.913 & 50/198 & 0  \\
0.3209 & 1.940 & 40/135 & 0  \\
0.2932 & 1.984 & 30/63.1 & 1  \\
0.2629 & 2.024 & 20/22.0 & 1  \\
0.2581 & 2.030 & 19/18.5 & 1  \\
0.2532 & 2.036 & 18/15.3 & 1  \\
0.2482 & 2.042 & 17/12.6 & 1  \\
0.2432 & 2.048 & 16/10.3 & 1  \\
0.2378 & 2.054 & 15/8.60 & 1  \\
0.2289 & 2.063 & 14/7.49 & 2  \\
0.2224 & 2.070 & 13/6.18 & 2  \\
0.2160 & 2.076 & 12/5.04 & 2  \\
0.2111 & 2.081 & 11/4.05 & 0  \\
0.2062 & 2.085 & 10/3.25 & 0  \\
0.2014 & 2.090 & 9/2.60 & 0  \\
0.1969 & 2.093 & 8/2.10 & 0 & $ 1.10 \times 10^{-5} $ & $ 9.20 \times 10^{-24} $ & $ 3.43 \times 10^{-22} $ & $ 1.09 \times 10^{6} $ \\
0.1925 & 2.097 & 7/1.70 & 0 & $ 1.36 \times 10^{-5} $ & $ 1.04 \times 10^{-23} $ & $ 4.30 \times 10^{-22} $ & $ 6.33 \times 10^{5} $ \\
0.1882 & 2.101 & 6/1.37 & 0 & $ 1.69 \times 10^{-5} $ & $ 1.17 \times 10^{-23} $ & $ 5.42 \times 10^{-22} $ & $ 3.63 \times 10^{5} $ \\
0.1839 & 2.104 & 5/1.10 & 0 & $ 2.10 \times 10^{-5} $ & $ 1.33 \times 10^{-23} $ & $ 6.85 \times 10^{-22} $ & $ 2.07 \times 10^{5} $ \\
0.1790 & 2.108 & 4/0.870 & 0 & $ 2.66 \times 10^{-5} $ & $ 1.52 \times 10^{-23} $ & $ 8.79 \times 10^{-22} $ & $ 1.13 \times 10^{5} $ \\
0.1730 & 2.112 & 3/0.636 & 0 & $ 3.64 \times 10^{-5} $ & $ 1.81 \times 10^{-23} $ & $ 1.23 \times 10^{-21} $ & $ 5.08 \times 10^{4} $ \\
0.1609 & 2.121 & 2/0.278 & 0 & $ 8.33 \times 10^{-5} $ & $ 2.94 \times 10^{-23} $ & $ 3.01 \times 10^{-21} $ & $ 5.98 \times 10^{3} $ \\
\hline
	\end{tabular}
\end{minipage}
\end{table*}
%

\begin{table*}
\begin{minipage}{140mm}
	\centering
	\caption{Grid of ELM WDs models considering the use of the enhanced CARB magnetic braking prescription proposed by \citet{Van2019}, as in Table~\ref{tab:M1014Z02-rot} but for $\beta_\text{mt}=0.8$, element diffusion and rotation is taking into account in all models.}
	\label{tab:M1014Z02-dif}
	\begin{tabular}{cccccccc}
\hline
$M_{\text{d,f}}~(M_{\odot})$ &
$M_{\text{a,f}}~(M_{\odot})$ &
$P_\text{i} / P_\text{f}~(\text{d})$ &
\#HSF &
$f_\text{GW}~\text{(Hz)}$ &
$\mathcal{A}$ &
$h_\text{c}$ &
$\tau_\text{GW}~\text{(Myr)}$ \\
\hline
0.4102 & 1.451 & 300/847 & 1\\
0.3936 & 1.465 & 200/648 & 1\\
0.3648 & 1.486 & 100/386 & 1\\
0.3492 & 1.496 & 70/277	& 1\\
0.3312 & 1.507 & 50/183	& 1\\
0.3165 & 1.515 & 40/124	& 1\\
0.2876 & 1.528 & 30/56.2 & 2\\
0.2706 & 1.535 & 25/31.3 & 2\\
0.2550 & 1.541 & 20/17.6 & 2\\
0.2502 & 1.542 & 19/14.7 & 2\\
0.2454 & 1.544 & 18/12.1 & 2\\
0.2405 & 1.545 & 17/10.0 & 2\\
0.2341 & 1.548 & 16/8.60 & 2\\
0.2219 & 1.551 & 15/7.73 & 3\\
0.2151 & 1.553 & 14/6.31 & 3\\
0.2082 & 1.555 & 13/5.09 & 3\\
0.2020 & 1.556 & 12/4.06 & 3\\
0.2967 & 1.558 & 11/3.19 & 3\\
0.1914 & 1.559 & 10/2.53 & 3\\
0.1852 & 1.560 & 9/2.03	& 3 & $1.14 \times 10^{-5}$ & $7.22 \times 10^{-24}$ & $2.74 \times 10^{-22}$ & $1.30 \times 10^{6}$\\
0.1809 & 1.561 & 8/1.59	& 3 & $1.46 \times 10^{-5}$ & $8.31 \times 10^{-24}$ & $3.56 \times 10^{-22}$ & $6.91 \times 10^{5}$\\
0.1698 & 1.564 & 7/1.40	& 4 & $1.65 \times 10^{-5}$ & $8.53 \times 10^{-24}$ & $3.89 \times 10^{-22}$ & $5.23 \times 10^{5}$\\
0.1697 & 1.564 & 6/1.00	& 3 & $2.31 \times 10^{-5}$ & $1.07 \times 10^{-23}$ & $5.76 \times 10^{-22}$ & $2.13 \times 10^{5}$\\
0.1586 & 1.566 & 5/0.815 & 3 & $2.84 \times 10^{-5}$ & $1.15 \times 10^{-23}$ & $6.86 \times 10^{-22}$ & $1.32 \times 10^{5}$\\
0.1504 & 1.568 & 4/0.489 & 2 & $4.73 \times 10^{-5}$ & $1.53 \times 10^{-23}$ & $1.18 \times 10^{-21}$ & $3.55 \times 10^{4}$\\
\hline
	\end{tabular}
\end{minipage}
\end{table*}
%

\section{Polynomial fits}
\label{sec:Appendix-PolFits}

In Table~\ref{tab:pol-fits} we show polynomial fits to the final ELM WD mass ($M_\text{d,f}$) as a function of the initial orbital period ($P_\text{i}$) for our models in the form of $y=\text{A}+\text{B}x+\text{C}x^{2}+\text{D}x^{3}+\text{E}x^{4}$.
The quality of the fits is indicated by the coefficient of determination ($R^{2}$).

\begin{table*}
\begin{minipage}{140mm}
	\centering
	\caption{Coefficients for the third degree polynomials that best fit the final ELM WD mass ($M_\text{d,f}$) as a function of the initial orbital period ($P_\text{i}$), i.e., $M_\text{d,f}$($P_\text{i}$).
	The fits are only valid for $M_\text{d,f} < 0.35~M_{\odot}$.
	The first column shows the setup, as in Table~\ref{tab:setups}. Columns two to six shows the coefficients in the form $y=\text{A}+\text{B}x+\text{C}x^{2}+\text{D}x^{3}+\text{E}x^{4}$, as discussed in the text.
	The last column shows the coefficient of determination ($R^{2}$).}
	\label{tab:pol-fits}
	\begin{tabular}{ccccccc}
		\hline
		\# & A & B & C & D & E & $R^{2}$\\
		\hline
		1 & $1.3209\times10^{-1}$ & $7.9247\times10^{-3}$ & $- 1.0617\times10^{-4}$ & $6.2084\times10^{-7}$ & $- 1.2798\times10^{-9}$ & 0.9977\\
		2 & $1.4805\times10^{-1}$ & $7.0866\times10^{-3}$ & $- 8.8185\times10^{-5}$ & $4.9107\times10^{-7}$ & $- 9.7816\times10^{-10}$ & 0.9980\\
		3 & $1.1415\times10^{-1}$ & $9.2348\times10^{-3}$ & $- 1.3749\times10^{-4}$ & $9.0360\times10^{-7}$ & $- 2.0138\times10^{-9}$ & 0.9972\\
		\hline
	\end{tabular}
\end{minipage}
\end{table*}
%

\section{Gravitational waves}
\label{sec:Appendix-GW}

The LISA sensitivity band frequency is from $10^{-1}$ to $10^{-5}$ Hz, with a peak around $4~\text{mHz}$. We can approximate the frequency of the gravitational wave emitted by the binary systems as $f_\text{GW}=2/P_\text{f}$, indicating that systems of ELM WDs with lower mass will populate the region of greatest sensitivity.

In order to estimate the gravitational radiation emitted by our models,
we adopt the same approach as in \citet{Kupfer2018,LiChenChenLiYuHan2020,Korol2020}.
We choose to present our estimates in terms of the characteristic strain since the difference between the source signal and the sensitivity of the detector is directly related to the signal-to-noise ratio \citep{Moore2015}.
Thus, with the data available in Appendix~\ref{sec:Appendix-CompMod}, it is immediate to obtain estimates for other values of observation time, distance, and also to calculate the signal-to-noise ratio for specific configurations of the detectors.

The signal-to-noise ratio is directly proportional to the dimensionless gravitational wave amplitude, that after averaging over inclination, sky-location and amplitude reads
\begin{equation}
\mathcal{A}=2 \pi^{2/3} \frac{G^{5/3}}{c^{4}}\frac{\mathcal{M}^{5/3} f_\text{GW}^{2/3}}{d} \text{,}
	\label{eq:h_0}
\end{equation}
where $\mathcal{M}=(M_\text{d,f} M_\text{a,f})^{3/5}(M_\text{d,f}+M_\text{a,f})^{-1/5}$ is the chirp mass and $d$ is the distance of the binary system to the Sun.

For an inspiralling binary system that emits monochromatic radiation, the characteristic strain is
\begin{equation}
h_\text{c}=\sqrt{N_\text{cycle}}\mathcal{A} \text{,}
	\label{eq:h_c}
\end{equation}
where $N_\text{cycle}=f_\text{GW} T_\text{obs}$ give us the total orbital periods observed over the detector's operation time.
For the purpose of facilitating comparisons and conversions, in this work we fix the distances to $d=1~\text{kpc}$ and
adopt the nominal LISA mission duration time of $T_\text{obs}=4~\text{yr}$.

When compared to the equation (2) of \citet{Brown2020ELMVIII} in the case of inclination $i=90 \degr$, our calculated characteristic strain is 1.42 times larger; or, alternatively, equivalent to $i \approx 66 \degr$.

Finally, starting from equation (2) in \citet{Kraft1962}, we obtained the following expression for the merging time due to gravitational radiation\footnote{We would like to point out that the multiplicative constant that we found is slightly different from that found by \citet{Brown2020ELMVIII}.}
\begin{equation}
\tau_\text{GW}=47100 \mathcal{M}^{-5/3} P_\text{f}^{8/3} \; \text{Myr}
	\label{eq:tau} \text{,}
\end{equation}
where again the chirp mass and the orbital period should be given in solar masses and days, respectively.
Since the expression above accounts only for the angular momentum loss due to the gravitational radiation ($\dot{J}_\text{gr}$), the merging time should be greater in models that may have significant contributions from other processes of loss of angular momentum (see equation~(\ref{eq:Jdot})). In our models, this is the case when $P_\text{i} \gtrsim 25~\text{d}$, where $\dot{J}_\text{ml}$ is the term that has the greatest contribution at the end of evolution.


\bsp	
\label{lastpage}
\end{document}